\documentclass[a4paper,fleqn]{cas-dc}

\usepackage[authoryear,longnamesfirst]{natbib}
\usepackage{bm}
\usepackage{graphicx}
\usepackage{epstopdf}
\usepackage{subfigure}

\def\tsc#1{\csdef{#1}{\textsc{\lowercase{#1}}\xspace}}
\tsc{WGM}
\tsc{QE}

\begin{document}
\let\WriteBookmarks\relax
\def\floatpagepagefraction{1}
\def\textpagefraction{.001}

\shorttitle{}    

\shortauthors{S. Chowdhury et~al.}  

\title [mode = title]{Complete Synchronization and its Transition in Higher Harmonic Sakaguchi--Kuramoto Oscillators}  



%

\author[1,2]{Subhasish Chowdhury}



\ead{subhasishchowdhury6@gmail.com}



\affiliation[1]{organization={Advanced Design and Analysis Group (ADAG), CSIR-Central Mechanical Engineering Research Institute},
            city={Durgapur},
            postcode={713209}, 
            country={India}}
\affiliation[2]{organization={Academy of Scientific and Innovative Research (AcSIR)},
            city={Ghaziabad},
            postcode={201002}, 
            country={India}}            

\author[3,4]{Sangita Dutta}




\affiliation[3]{organization={Department of Mathematics, National Institute of Technology },
            city={Durgapur},
            postcode={713209}, 
            country={India}}
\affiliation[4]{organization={Institute of Technical Physics and Materials Science, HUN-REN Centre for Energy Research},
            city={Budapest},
            postcode={1525}, 
            country={Hungary}}
\author[5]{Pitambar Khanra}
\credit{}

\affiliation[5]{organization={School of Computer Science, UPES },
            city={Dehradun},
            postcode={248007}, 
            country={India}}
\author[1,2]{Swarup Kumar Laha}





\credit{} 
\author[6]{Prosenjit Kundu}

\ead{jitprosen.math@gmail.com}

\cormark[1]

\credit{}

\affiliation[6]{organization={CSys Lab, Dhirubhai Ambani University},
            city={Gandhinagar},
            postcode={382007}, 
            country={India}}

\cortext[1]{Corresponding author}



\begin{abstract}
In heterogeneous networks of coupled oscillators, phase frustration typically prevents the emergence of synchronization in the Sakaguchi-Kuramoto (SK) model. In this study, we propose an analytical framework to overcome this barrier and induce complete synchronization at a specified small coupling value in oscillators governed by phase-frustrated bi-harmonic coupling. We derive an optimal set of natural frequencies that is robust against added noise and  correlated with the network's degree heterogeneity, along with the parameters involved in the bi-harmonic coupling function that lead to complete synchronization $(r=1)$. On top of that, we found complete synchronization transition along with hysteresis in the case of scale-free networks, indicating a first-order (discontinuous) phase transition, whereas Erdős–Rényi networks exhibit complete synchronization transition with a second-order (continuous) phase transition. Furthermore, we use the mean-field approximation in the presence of optimal frequency to determine the critical coupling strength associated with the synchronization transition in the pure second-harmonic Sakaguchi-Kuramoto model. Here, the obtained optimal natural frequencies scale linearly with the node degree, and the critical coupling strength for the onset of synchronization is derived analytically from the self-consistent equations. In this specific regime, we observe a perfectly ordered two cluster synchronized state. These findings remain robust with higher-order harmonic coupling schemes, as well as across a diverse range of synthetic and empirical networks, including scale-free, Erdős–Rényi, Zachary Karate Club and \textit{C.~elegans} neural network, demonstrating their general applicability.
\end{abstract}




\begin{keywords}
Synchronization \sep Bi-harmonic coupling \sep Sakaguchi-Kuramoto model 
\end{keywords}
\maketitle 
\section{Introduction}
Dynamical units often exhibit diverse collective phenomena when connected through complex network topologies. One of the most prominent emergent behaviors is synchronization \cite{pikovsky2001synchronization, arenas2008synchronization}, a phenomenon observed and tested in various real-world fields. For instance, coherent behavior emerges in coupled cellular systems and power distribution grids, while both desired and undesired coherent bursts are common features in populations of neuronal activity \cite{Strogatz_synchronization_book, pikovsky2001synchronization, arenas2008synchronization, Motter_NatPhys2013, Belykh_PRL2005, Dorogovtsev_RMP2008}. Characterizing the statistical scaling of various complex network topologies has advanced significantly in the last few decades \cite{Dorogovtsev_RMP2008, Albert_RMP2002, Cohen_complex_book,kundu2020optimizing,jalan2017unveiling,bandyopadhyay2007universality}. Consequently, research has increasingly focused on uncovering the interplay between the structural heterogeneity of a network and the oscillatory or steady-state dynamics of its individual nodes \cite{Barzel_NatPhys2013}. In addition, the interlink between the complex structure of a network and the macroscopic patterns emerging from node dynamics, such as synchronization, is investigated with a profound analytical description for different cases, ranging from networks of chaotic dynamics \cite{arenas2008synchronization, ichinomiya2004frequency,Restrepo_PRE2005} to networks of higher-order interactions \cite{jain2025low,sharma2024synchronization,dutta2024transition}.

The pioneering works of Winfree \cite{winfree1967biological} and Kuramoto \cite{kuramoto1984chemical} established the theoretical foundation for these studies. The Kuramoto model, in particular, has become a central and extensively investigated paradigm for understanding collective synchronization in oscillator networks \cite{Rodrigues_PhysReport2016,Acebron_RMP2005}. In this framework, oscillators with distinct natural frequencies interact through pairwise couplings. Depending on the organization of coupling strengths or frequencies, these systems can display various forms of collective behavior, including continuous or abrupt (discontinuous) synchronization transitions \cite{kuramoto1984chemical,gomez2011explosive,gomez2007paths,leyva2012explosive,coutinho2013kuramoto,ichinomiya2004frequency}. An important generalization of this framework is the Sakaguchi-Kuramoto model, which incorporates a phase-lag parameter into the interaction term \cite{sakaguchi1986soluble}. This phase lag accounts for interaction delays common in natural and engineered systems. However, the introduction of even a small phase shift significantly influences stability, often completely suppressing abrupt synchronization transitions \cite{lohe2015synchronization,kundu2018perfect,kundu2020optimizing,kundu2017transition,skardal2015erosion}. This leaves a critical open question: can a specific configuration of intrinsic parameters force a frustrated, bi-harmonic system into a state of complete synchrony at some targeted coupling value?

While synchronization under single-harmonic coupling has been extensively investigated \cite{kuramoto1975formation,Strogatz_PhyD2000,kundu2018perfect,kundu2017transition}, recent years have seen a surge of interest in bi-harmonic Kuramoto models, which offer a richer framework for complex dynamics \cite{komarov2013multiplicity,komarov2014kuramoto,li2014transition,wang2017synchronous,eydam2017mode,lohe2015synchronization}. These higher-order coupling terms are not merely theoretical abstractions, but are intrinsic to several physical systems. A classical example is the Huygens' clock setup, where the horizontal motion of the supporting beam induces a first harmonic interaction, while vertical vibrations contribute a second harmonic component \cite{czolczynski2013synchronization,komarov2014kuramoto}. Similarly, in Josephson junctions, phase dynamics evolving within a tilted double-well potential naturally enhance second harmonic contributions \cite{goldobin2011josephson,goldobin2013phase}. On the other hand, second harmonic terms have also been identified in the coupling functions of globally coupled electrochemical oscillators \cite{kiss2005predicting,kiss2006characteristics}.

Despite the theoretical richness of the Sakaguchi-Kuramoto model with bi-harmonic coupling, several fundamental aspects remain underexplored. Notably, complete synchronization is generally reported to be unattainable in these phase-frustrated networks, regardless of coupling strength. Furthermore, the specific role of the phase frustrated second harmonic term and the potential for synchronization in the absence of the first harmonic remain insufficiently understood.

In this paper, we propose a scheme to bridge these gaps by deriving a specific set of optimal natural frequencies that allows the system to achieve complete synchronization, even in the presence of phase frustration. Unlike previous works, we analytically derive this frequency set based on the network's degree heterogeneity and the specific phase-frustrated coupling function.

The paper is organized as follows: In Section \ref{Mathematical Model} , we introduce the mathematical model. In Section \ref{Analytical Derivation}, we perform the analytical derivation to find the optimal frequency set. We also mathematically explore the effect of multiplicative noise in the natural frequency set on synchronization. Further, we employ a mean-field approach to analytically determine the critical coupling strength of the pure second-order SK model. In Section \ref{Results}, we present numerical results that demonstrate complete synchronization for the bi-harmonic SK model and complete two cluster synchronization for the pure second-harmonic SK model at some targeted coupling parameter. Finally, in the Appendix, we demonstrate the generalization of our framework .
\begin{figure*}[t]
\centering
\includegraphics[width=\textwidth]{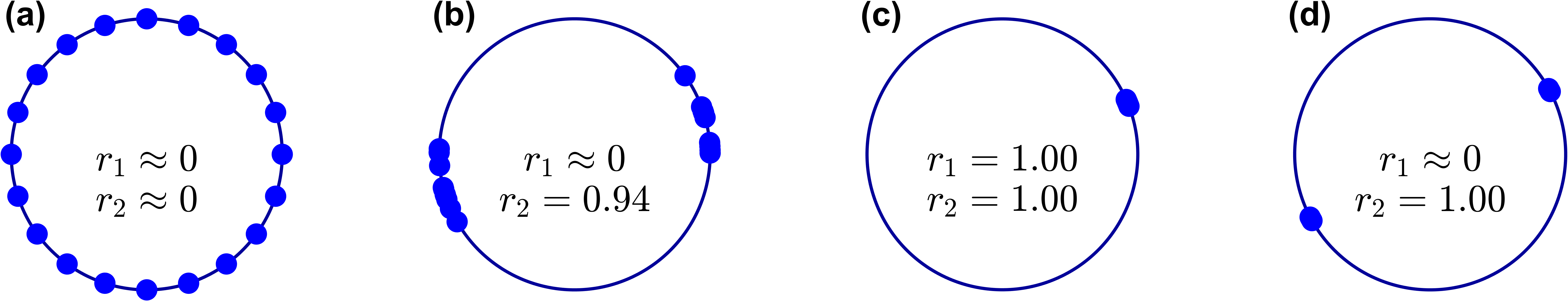}

\caption{
\textbf{Different Synchronization states for different values of order parameters.} 
(a) Incoherent state. 
(b) Symmetric two-cluster synchronized state. 
(c) Complete single-cluster synchronized state. 
(d) Complete symmetric two-cluster synchronized state.
\cite{skardal2011cluster}}
\label{fig:cluster_states}
\end{figure*}

\section{Mathematical Model}\label{Mathematical Model}
 We consider the bi-harmonic SK model of $N$ coupled oscillators (nodes), whose phases follow the general dynamics
\cite{sakaguchi1986soluble,Kuramoto_chemical_book}
 \begin{eqnarray}
\frac{d\theta_i}{dt} &=& \omega_i + {K_1}\sum_{j=1}^{N} A_{ij}{F}(\theta_j - \theta_i-\alpha_{ij}^{(1)}) \nonumber \\ &+& {K_2}\sum_{j=1}^{N} A_{ij}{F}(2\theta_j - 2\theta_i-\alpha_{ij}^{(2)})
\label{eq:General_bi_harmonic}
\end{eqnarray}
where, $\omega_i$ denotes the natural frequency of node $i$, $A=(A_{ij})_{N \times N}$ is an adjacency matrix that determines the underlying interaction network,
and the parameters
$K_1$ and $K_2$ are the coupling strengths (or values) for the first and second harmonic functions, respectively.
The functional forms of the coupling are represented by 
$F(\theta_j - \theta_i - \alpha^{(1)}_{ij})$ and 
$F(2\theta_j - 2\theta_i - \alpha^{(2)}_{ij})$, 
which are $2\pi$-periodic functions. The distributed phase lags 
$\alpha^{(1)}_{ij}$ and $\alpha^{(2)}_{ij}$ characterize the presence 
of frustration in the respective coupling interactions.

We would like to mention here 
$F(\theta_j - \theta_i - \alpha^{(1)}_{ij}) = \sin(\theta_j - \theta_i - \alpha)$ and $F(2\theta_j - 2\theta_i - \alpha^{(2)}_{ij}) = \sin(2\theta_j - 2\theta_i - \beta)$ (when all $\alpha_{ij}^{(1)}=\alpha$ and $\alpha_{ij}^{(2)}=\beta$ for all $i,j$),
Eq. (\ref{eq:General_bi_harmonic}) represents the Sakaguchi-Kuramoto model with a bi-harmonic coupling function, and in the limit 
$\alpha \rightarrow 0$ and $\beta \rightarrow 0$
 converges to the classic Kuramoto dynamics with a bi-harmonic coupling function. 
To quantify synchronization in the coupled network, we use the
Kuramoto order parameters
\begin{equation*}
r_m = \left| \frac{1}{N} \sum_{j=1}^{N} e^{i m \theta_j} \right|,
\quad m \in \mathbb{N}.
\end{equation*}
The quantity $r_m \approx 0$ corresponds to the incoherent state,
whereas $r_m > 0$ indicates the presence of phase coherence. Complete synchronization, defined by
$\theta_1 = \theta_2 = \cdots = \theta_N$, yields $r_1 = 1$ \cite{kundu2018perfect} and $r_1 \rightarrow 0 $ , $r_2=1$ simultaneously corresponds to a complete symmetric two-cluster synchronized state \cite{skardal2011cluster}.
Synchronization depends on the interplay between the spread of
natural frequencies $\omega_i$ and the coupling strength
$K$~\cite{kuramoto1975formation}. Thus, coherence can be achieved
either by reducing the frequency dispersion (e.g., drawing
$\omega_i$ from a narrow normal distribution or from a homogeneous distribution) or by increasing
the coupling strengths $K_1$ and $K_2$ until order parameter reaches unity.

The difficulty arises because, in the presence of phase frustration 
$(\alpha, \beta > 0)$, complete synchronization cannot be achieved. 
This remains true even when the coupling strengths $K_1$ and $K_2$ 
are taken to be very large or when the natural frequencies $\omega_i$ 
are highly homogeneous. Indeed, a small amount of lag is sufficient in order to prohibit
the system from reaching complete synchrony. 
 Contrast to all previous works,  we show  here that by strategically assigning all nodes with a specific set of natural frequencies,the system Eq. \eqref{eq:General_bi_harmonic}  can be able to drive itself into complete synchronization  ($r \sim1$) with even small coupling strength. 
We derive such a specific set of {\it optimal frequency sequences},
$\boldsymbol{\omega} = \{\omega_i\}_{i = 1}^N$, based on network structure $A_{ij}$, coupling strength
and the activity of the phase frustrated coupling function $(F)$.

\section{Analytical Derivation}\label{Analytical Derivation}
\subsection{Optimal frequency for complete synchronization}\label{Optimal frequency for complete synchronization}
To obtain 
$\boldsymbol{\omega}$,
we analyze Eq. \eqref{eq:General_bi_harmonic} as it approaches synchronization \cite{skardal2014optimal,Skardal_PRE2014,Skardal_PhysicaD2014}, that is, in the limit 
$|\theta_j - \theta_i| \rightarrow 0$.
In this limit, the coupling functions can be approximated by
$F(\theta_j - \theta_i - \alpha_{ij}^{(1)}) \approx F(-\alpha_{ij}^{(1)}) + F^{\prime}(-\alpha_{ij}^{(1)})(\theta_j - \theta_i) + \dots$,

$F(2\theta_j - 2\theta_i - \alpha_{ij}^{(2)}) \approx F(-\alpha_{ij}^{(2)}) + 2F^{\prime}(-\alpha_{ij}^{(2)})(\theta_j - \theta_i) + \dots$,
where
$F^{\prime}(\alpha) = \left. \frac{dF(\theta, \alpha)}{d\theta} \right|_{\theta = 0}$. This allows us to write Eq. \eqref{eq:General_bi_harmonic} in its linearized form as
\begin{align}
\frac{d\theta_i}{dt} 
&= \omega_i 
+ \sum_{j=1}^{N} A_{ij} \Big[ K_{1}F(-\alpha_{ij}^{(1)}) 
   + K_{2}F(-\alpha_{ij}^{(2)}) \Big] \nonumber \\
& - \sum_{j=1}^{N} \Bigg[
   \delta_{ij} \Bigg( \sum_{l=1}^{N} A_{il} 
   \Big[ K_{1}F^{\prime}(-\alpha_{il}^{(1)}) 
   + 2K_{2}F^{\prime}(-\alpha_{il}^{(2)}) \Big] \Bigg) \nonumber \\
&\ - A_{ij} \Big[ K_{1}F^{\prime}(-\alpha_{ij}^{(1)}) 
   + 2K_{2}F^{\prime}(-\alpha_{ij}^{(2)}) \Big] \Bigg] \theta_j
\label{eqn2}
\end{align}
or in vector form,
\begin{eqnarray}
\frac{d{\boldsymbol\theta}}{dt}={\boldsymbol \omega}+  {\boldsymbol  {\tilde{d}}} - \tilde{\boldsymbol L} {\boldsymbol \theta}
\label{eqn3(a)}
\end{eqnarray}
\noindent
Here $\tilde{\boldsymbol L}$ and ${\boldsymbol {\tilde{d}}}$ are the Laplacian matrix and degree vector, respectively, which can be written as 
$$\tilde{d}_i = \sum_{j=1}^{N} A_{ij} \big[K_{1}F(-\alpha_{ij}^{(1)})+K_{2}F(-\alpha_{ij}^{(2)})\big]$$
and 
\begin{align}
\tilde{L}_{ij} &= \Bigg(\sum_{l=1}^{N} A_{il}\Big[K_{1}F^{\prime}(-\alpha_{il}^{(1)}) + 2K_{2}F^{\prime}(-\alpha_{il}^{(2)})\Big]\Bigg) \nonumber \\
&\quad - A_{ij} \Big[K_{1}F^{\prime}(-\alpha_{ij}^{(1)}) + 2K_{2}F^{\prime}(-\alpha_{ij}^{(2)})\Big] \nonumber
\end{align}
The system will reach a synchronized state if, for some choice of the natural frequencies 
$\omega_i$,
Eq.\ (\ref{eqn3(a)}) gets a solution where all phases,
$\theta_i$,
change according to some global frequency
$\Omega$ 
as
$\theta_i(t) = \Omega t$,
independent of $i$.
So, steady state form, in the rotating frame
\begin{eqnarray}
\boldsymbol{\tilde{L}} {\boldsymbol {\theta^*}}=( {{\boldsymbol{\omega}}+  {\boldsymbol{\tilde{d}}}-\boldsymbol{{\Omega}}} )
\label{eqn4}
\end{eqnarray}
Here, vector 
${\boldsymbol {\theta^*}}$ 
is the steady state solutions of the full network in  the rotating frame. Finally, the steady states look like
\begin{equation}
{\boldsymbol {\theta^*}} = \boldsymbol{\tilde{L^{\dagger}}}  {({\boldsymbol{\omega}} +   {\boldsymbol{\tilde{d}}}-{\boldsymbol{\Omega}})}
\label{fp1}
\end{equation} 
Here, $\boldsymbol{\tilde{L}^{\dagger}}$ denotes the pseudo-inverse of the Laplacian operator \cite{ben2003generalized}. 
Since the Laplacian matrix is symmetric and positive semi-definite, its eigenvectors $\boldsymbol{v}_i$ form an orthonormal basis of $\mathbb{R}^N$, and the corresponding eigenvalues $\lambda_i$ are real and non-negative, ordered as
\begin{equation}
0 = \lambda_1 < \lambda_2 \le \cdots \le \lambda_N .
\end{equation}

Using the spectral decomposition \cite{sarkar2018spectral}   of the Laplacian, the pseudo-inverse operator can be written as
\begin{equation}
\boldsymbol{\tilde{L}^{\dagger}}
= \sum_{j=2}^{N} \lambda_j^{-1} \, \boldsymbol{v}_j \boldsymbol{v}_j^{T},
\end{equation}

Under the small-phase approximation $ \lVert \boldsymbol{\theta}^* \rVert \ll 1 $, the $m$-th order parameter can be approximated as
\begin{equation}
r_m \approx 1 - \frac{\lVert m \boldsymbol{\theta}^* \rVert^2}{2N},
\end{equation}
which quantifies the deviation from complete synchronization in terms of the squared Euclidean norm of the steady-state phase vector.

 Now, 
 \begin{align*}
\|m\bm{\theta}^{*}\|^{2}
  &= \langle m\bm{\theta}^{*},\, m\bm{\theta}^{*} \rangle \\
  &= m^{2}\Big\langle \bm{\tilde{L}}^{\dagger}(\bm{\omega}+\bm{\tilde d}),
                   \,\bm{\tilde{L}}^{\dagger}(\bm{\omega}+\bm{\tilde d})\Big\rangle \\
  &= m^{2}\sum_{j=2}^{N}\lambda_j^{-2}
     \,\big\langle v_j,\,(\bm{\omega}+\bm{\tilde d})\big\rangle^{2}.
\end{align*}
Therefore,
\begin{equation}
    r_{m}
   = 1 - \frac{m^{2}}{2N}
     \sum_{j=2}^{N}\lambda_{j}^{-2}
     \big\langle v_{j},\,(\boldsymbol{\omega}+\boldsymbol{\tilde d})\big\rangle^{2}
   = 1 - \frac{m^{2}}{2}\,J(\boldsymbol{\omega},\boldsymbol{L}),
   \label{eq:r_m}
\end{equation}
where $J(\boldsymbol{\omega},\boldsymbol{L})$ denotes the synchrony alignment function (SAF),
\[
J(\bm{\omega},\bm{L})
   = \frac{1}{N}
     \sum_{j=2}^{N}\lambda_{j}^{-2}
     \big\langle v_{j},\,(\bm{\omega}+\bm{\tilde d})\big\rangle^{2}.
\]
For complete synchronization state, we have $J(\bm{\omega},\bm{L})\to 0$, Eq. \eqref{eq:r_m} gives $r_m=1$ ( complete m-th order synchronization or perfect m-cluster coherence \cite{skardal2011cluster}). Now, we can find out {\it optimal frequency sequence},
${\boldsymbol { \omega}} = \{\omega_i\}_{i = 1}^N$ by taking,
\begin{equation}
\begin{split}
J(\bm{\omega},\bm{L})=0 
&\;\implies\; \bm{\omega}+\bm{\tilde d} = 0 \\
&\;\implies\;
\boldsymbol{\omega} = -\sum_{j=1}^{N} A_{ij}
      \Big[\,K_{1}F(-\alpha_{ij}^{(1)})  \\
&\hspace{4.6em}
      +\,K_{2}F(-\alpha_{ij}^{(2)})\,\Big].
\end{split}
\label{eq:J(w,L)}
\end{equation}

In our case,
\begin{equation}
    \bm{\omega}= \sum_{j=1}^{N} A_{ij} [K_{1}Sin(\alpha)+K_{2}Sin(\beta)]
    \label{eq:optimal_frequency}
\end{equation}

In this rotating frame, it is obvious that   the stability of these fixed points (${\boldsymbol {\theta^*}}$) 
will ensure the stability of the synchrony in the heterogeneous network of phase frustrated oscillators. These fixed  points depend on several components (see Eq. \ref{fp1}): coupling strength $(K)$, the desired frequency set ($\bm{\omega} = \{\omega_i\}_{i = 1}^N$), and degree vector (${\boldsymbol{\tilde{d}}}$) of the interacting network. 
Here we observe that the analytically derived frequency set depends on the degree distribution of the underlying network. 
Such degree-frequency correlations are known to induce abrupt or explosive synchronization transitions in some scale-free networks \cite{gomez2011explosive,Coutinho_PRE2013}.

\subsection{Synchronization under the Presence of Noise}\label{Synchronization under the Presence of Noise}
Examine the response of synchronization to variations in the optimal frequency set (\ref{eq:optimal_frequency}) by adding Gaussian noise $\boldsymbol{\delta \omega}$ to the optimal frequency $\boldsymbol{\omega}$ such that ${\omega_i} \rightarrow \omega_i + \delta\omega_i$, where $\delta \omega_i \sim \mathcal{N}(0,\sigma{\omega_i})$, that is, a normal random variable with mean $0$ and variance $\sigma^2{\omega_i}^2$, acting as a multiplicative perturbation of $\omega_i$. This noise can decrease the level of synchronization from a complete state to a partial state or de-synchronized state. For this measurement, let us consider a function, $\eta = 1 - r_m,$ where $\quad 0 \leq \eta  \leq 1 $ and $r_{m} \approx 1 - {\lVert{\boldsymbol {m\theta}}\rVert}^2/{2N}$ \\ Therefore,
\begin{align}
\eta &= 1 - r_m \nonumber \\
       &\approx 1 - \left( 1 - \frac{\lVert \boldsymbol{m\theta} \rVert^2}{2N} \right) \nonumber \\
       &=\frac{m^2}{2}{Var}\left(\bm{\theta} \right)
       \label{eq:eta}
\end{align}
 After adding noise $\boldsymbol{\delta \omega}$, from (\ref{fp1}) we get
\begin{align*}
    { {\theta_i}} &= \sum_{j}{\tilde{L_{ij}^{\dagger}}}  {({{\omega_j}} + {\delta \omega_j} +  {{\tilde{d_j}}}-{{\Omega}})}\\
    &=\sum_{j} {\tilde{L_{ij}^{\dagger}}} ({\delta \omega_j})
\end{align*}
Now,
\begin{align}
    \frac{m^2}{2} {Var}(\bm{\theta})
    &= \frac{m^2}{2}\sum_{j} \bigl(\tilde{L}_{ij}^{\dagger}\bigr)^{2} \,{Var}\bigl(\delta \omega_{j}\bigr) \nonumber \\[4pt]
    &\approx \frac{m^2}{2}\sum_{j} \bigl(\tilde{L}_{ij}^{\dagger}\bigr)^{2} \,\omega_{j}^{2}\,\sigma^{2} \nonumber \\[4pt]
    &= Z_{i}\,\sigma^{2};
     \text{where},\, Z_{i}=\frac{m^2}{2}\sum_{j} \bigl(\tilde{L}_{ij}^{\dagger}\bigr)^{2}\,\omega_{j}^{2}.
    \label{eq:Var}
\end{align}
Therefore, from \eqref{eq:eta} and \eqref{eq:Var},  
\begin{align}
 \eta\sim  \sigma^2 
 \label{eq:noise}
\end{align}
The scaling relation $\eta \sim \sigma^2$ indicates that the loss of
synchronization depends quadratically on the noise strength in the
oscillator frequencies. This provides a quantitative measure of the
degradation of synchronization as the natural frequencies deviate from
the optimal configuration given in Eq.~\eqref{eq:optimal_frequency}. In the weak-noise
limit, the behavior follows the theoretical prediction in
Eq.~\eqref{eq:noise}. As $\sigma$ increases, the natural frequencies
progressively depart from the optimal set defined by
Eq.~\eqref{eq:optimal_frequency}, eventually leading to a complete loss of
synchronization.

\subsection{Mean field Analysis} \label{Mean field Analysis}
 In the absence of the first harmonic coupling strength, i.e., when $K_{1}=0$, the interaction between the oscillators is governed solely by the second harmonic term \cite{skardal2011cluster}. This model looks similar in form to the general SK model \cite{sakaguchi1986soluble} but not equivalent since, second harmonic SK model can not be transform into general SK model and vice versa \cite{li2019clustering}. 
\begin{equation}
      \frac{d\theta_i}{dt} = \omega_i + {K_2}\sum_{j=1}^{N} A_{ij}\sin(2\theta_j - 2\theta_i-\beta)
      \label{eq:2nd_harmonic}
\end{equation}
Here, the natural frequency $\omega_i$ is selected from the optimal frequency set that allows the system to achieve complete symmetric two-cluster synchronized state at the targeted coupling strength $K_2=K_2^{(p)}$, as defined by Eq.~(\ref{eq:optimal_frequency})
\begin{equation}
\omega_i = K_2^{(p)} \sum_{j=1}^{N} A_{ij}\sin\beta = a\, q_i
\label{eq:optimal_K}
\end{equation}
where $a = K^{opt}_2 \sin\beta$ and $q_i = \sum_{j=1}^{N} A_{ij}$ denotes the degree of the $i$-th node.
 
 To analyze the dynamics, we apply a mean-field framework to the above Eq. (\ref{eq:2nd_harmonic}), following the procedure introduced in \cite{ichinomiya2004frequency,komarov2013multiplicity}. Let $\rho(q,\theta,t)$ represent the density of the nodes of degree $q$ having phase $\theta$ at time $t$. This density is normalized over the phase variable such that
\begin{equation}
\int_{0}^{2\pi} \rho(q,\theta,t)\, d\theta = 1.
\end{equation}
Assuming that the network does not exhibit degree--degree correlations, the probability that a randomly selected edge is connected to a node of degree $q$ with phase $\theta$ at time $t$ is given by
\[
\frac{qP(q)\,\rho(q,\theta,t)}{\int qP(q)\,dq},
\]
where $P(q)$ is the degree distribution of the network.

In the continuum limit, Eq. (\ref{eq:2nd_harmonic}) can be expressed as  
\begin{equation}
    \frac{d\theta}{dt} =\omega 
    + \frac{K_2q}{\langle q \rangle} \int dq' \int q' P(q') \rho(q', \theta', t) 
    \sin(2\theta' - 2\theta - \beta) \, d\theta',
    \label{eq:continum_equation}
\end{equation}
where the quantity $\langle q \rangle = \int qP(q)\,dq$ represents the mean degree of the network.To ensure the conservation of oscillators in Eq. (\ref{eq:2nd_harmonic}), the density function $\rho$ must satisfy the continuity equation
\begin{equation}
\frac{\partial \rho}{\partial t} + \frac{\partial}{\partial \theta}(\rho v) = 0,
\label{eq:continuity}
\end{equation}
where $v$ is taken from the right-hand side of Eq. (\ref{eq:continum_equation}).

To quantify the macroscopic dynamics of the oscillator population, we consider the order parameter $r_2$ in the thermodynamic limit, following the formulation presented in~\cite{ichinomiya2004frequency}
\begin{equation}
    r_{2}e^{i\psi}=\frac{1}{\langle q \rangle}\int dq{'}\int q^{'}P(q{'})\rho(q{'},\theta{'},t)e^{i2\theta'}d\theta' ,
    \label{eq:order_parameter}
\end{equation}
where $\psi$ denotes the mean phase of the oscillators, and the order parameter $r_2$ lies in the interval $0 \le r_{2} \le 1$.

Using the order parameter Eq. (\ref{eq:order_parameter}) , Eq. (\ref{eq:continum_equation}) can be rewritten as
\begin{equation}
    \frac{d\theta}{dt}=\omega+K_2r_2qsin(\psi-2\theta-\beta)
    \label{eq:Governing}
\end{equation}

To derive the self-consistent equations, we fix the global phase as $\psi(t)=\Omega t$, where $\Omega$ denotes the collective angular velocity, and introduce a new phase variable $\chi$ defined by
\[
\chi(t) = 2\theta(t) - \psi(t) + \beta .
\]
In this rotating frame, Eq. (\ref{eq:Governing}) can be rewritten as
\begin{equation}
    \frac{d\chi}{dt}=2aq-\Omega-2qK_{2}r_{2}sin(\chi)
    \label{eq:final_Gov}
\end{equation}
Then continuity Eq. (\ref{eq:continuity}) can be written in the form
\begin{equation}
    \frac{\partial}{\partial t}\rho(q,\chi,t)+\frac{\partial}{\partial \chi}\!\left[v_{\chi}\,\rho(q,\chi,t)\right]=0 , \text{where}, v_{\chi} = \frac{d\chi}{dt}
\end{equation}

In the steady state, $\frac{\partial}{\partial t}\rho(q,\chi,t)=0$, and hence the steady-state solution for the density function $\rho$ as follows

\begin{equation}
    \rho(q, \chi) = 
\begin{cases}
\delta\left( \chi - \arcsin\left( \frac{2aq - \Omega}{2qK_2 r_2} \right) \right), & \left| \frac{2aq - \Omega}{2qK_2r_2 } \right| \leq 1, \\
\frac{A(q)}{2aq - \Omega - 2qK_2r_2\sin(\chi)}, & \left| \frac{2aq - \Omega}{2qK_2r_2 } \right| > 1,
\end{cases}
\end{equation}
where $\delta$ is the Dirac delta function and $A(q)$ is the normalization constant given by $A(q)=\frac{\sqrt{(2aq-\Omega)^2-(2qK_2r_2)}}{2\pi}.$ The first solution represents the synchronous state, while the second solution is associated with the desynchronous state. Consequently, the order parameter $r_2$ can be rewritten as
\begin{align}
    r_2 &= \frac{1}{\langle q \rangle} \int d\chi \Biggl[
    \, \int_{q_{\min}}^{\infty} dq\, q P(q) \rho(q,\chi) e^{i(\chi - \beta)} \nonumber \\
    &\qquad \times H\left( 1 - \left| \frac{2aq - \Omega}{2q K_2 r_2} \right| \right) \nonumber \\
    &\quad + \int_{q_{\min}}^{\infty} dq\, q P(q) \rho(q,\chi) e^{i(\chi - \beta)} \nonumber \\
    &\qquad \times H\left( \left| \frac{2aq - \Omega}{2q K_2 r_2} \right| - 1 \right) \Biggr]
    \label{eq:order_para_r2}
\end{align}
where $H$ is heaviside function. Here, the first term on the right-hand side of Eq. (\ref{eq:order_para_r2}) arises from the contribution of the locked oscillators, while the second term corresponds to the contribution of the drifting oscillators to the order parameter~$r_2$.
 The contribution of locked oscillators to the order parameter is ,
\begin{align}
r_2^{\text{lock}}
&=
\Bigg[
\frac{\cos\beta}{\langle q \rangle}
\int_{q_{\min}}^\infty dq\, qP(q)
\sqrt{1-\!\left(\frac{2aq-\Omega}{2qK_2r_2}\right)^2}
+
\frac{\sin\beta}{\langle q \rangle}
 \nonumber \\   
&\quad\times
\int_{q_{\min}}^\infty dq\, qP(q)
\left(\frac{2aq-\Omega}{2qK_2r_2}\right)\Bigg]
H\!\left(1-\left|\frac{2aq-\Omega}{2qK_2r_2}\right|\right)
\nonumber \\          
&
-i
\Bigg[
\frac{\sin\beta}{\langle q \rangle}
\int_{q_{\min}}^\infty dq\, qP(q)
\sqrt{1-\!\left(\frac{2aq-\Omega}{2qK_2r_2}\right)^2}
-
\frac{\cos\beta}{\langle q \rangle}
 \nonumber \\   
&\quad\times
\int_{q_{\min}}^\infty dq\, qP(q)
\left(\frac{2aq-\Omega}{2qK_2r_2}\right)
\Bigg]H\!\left(1-\left|\frac{2aq-\Omega}{2qK_2r_2}\right|\right)
\label{eq:locked}
\end{align}
For drift oscillators ,
\begin{align}
r_2^{\text{drift}}
&=
\frac{\sin\beta + i\cos\beta}{\langle q \rangle}
\int_{q_{\min}}^\infty
dq\, P(q)
\left(\frac{2aq-\Omega}{2K_2 r_2}\right)
 \nonumber \\
&\quad\times \Bigg[
1-\sqrt{1-
\left(\frac{2qK_2 r_2}{2aq-\Omega}\right)^2}
\Bigg]
H\!\left(
\left|\frac{2aq-\Omega}{2qK_2 r_2}\right|-1
\right)
\label{eq:Drift}
\end{align}
\begin{figure*}[t]
\centering

\subfigure{\includegraphics[width=0.45\textwidth]{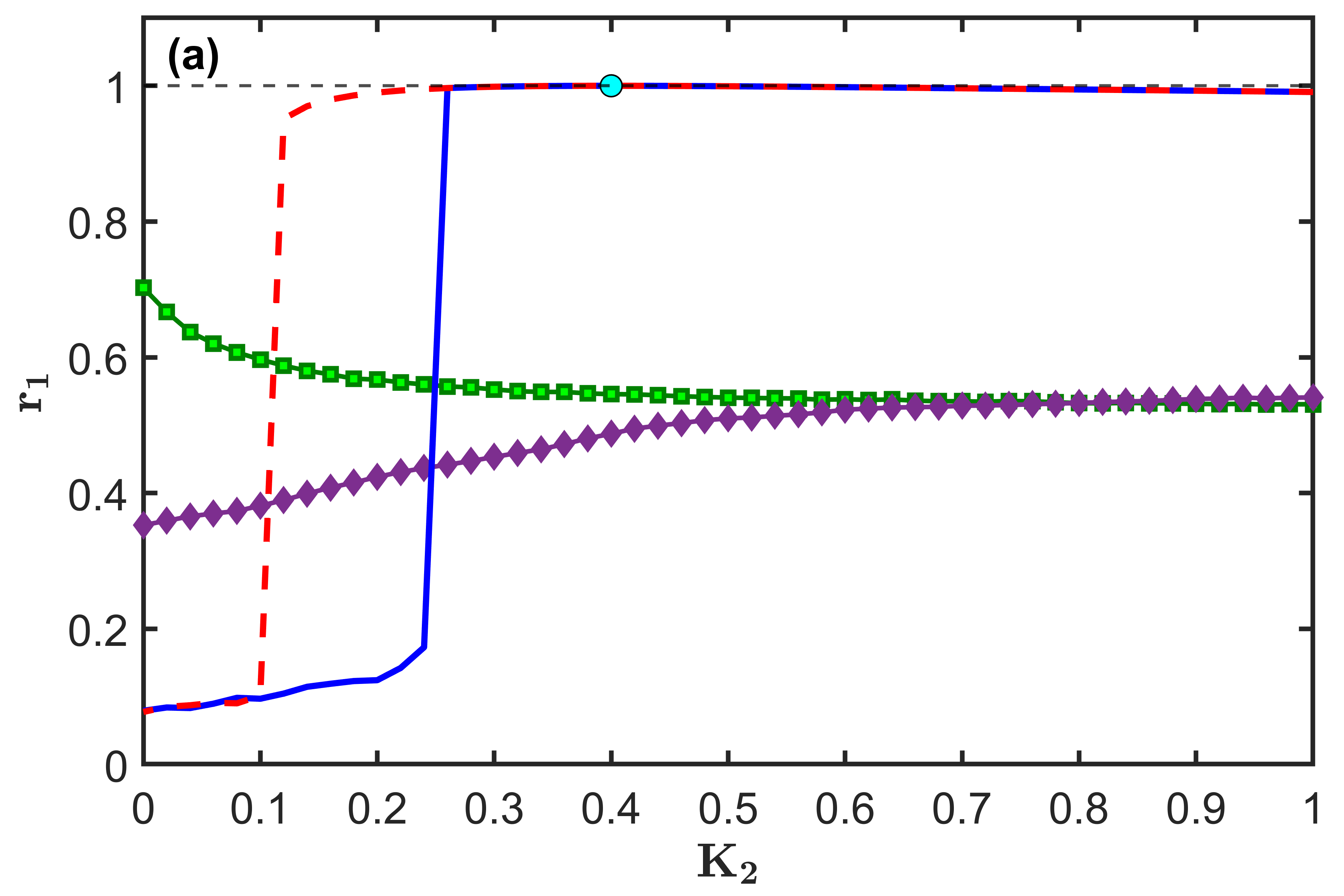}}
\subfigure{\includegraphics[width=0.45\textwidth]{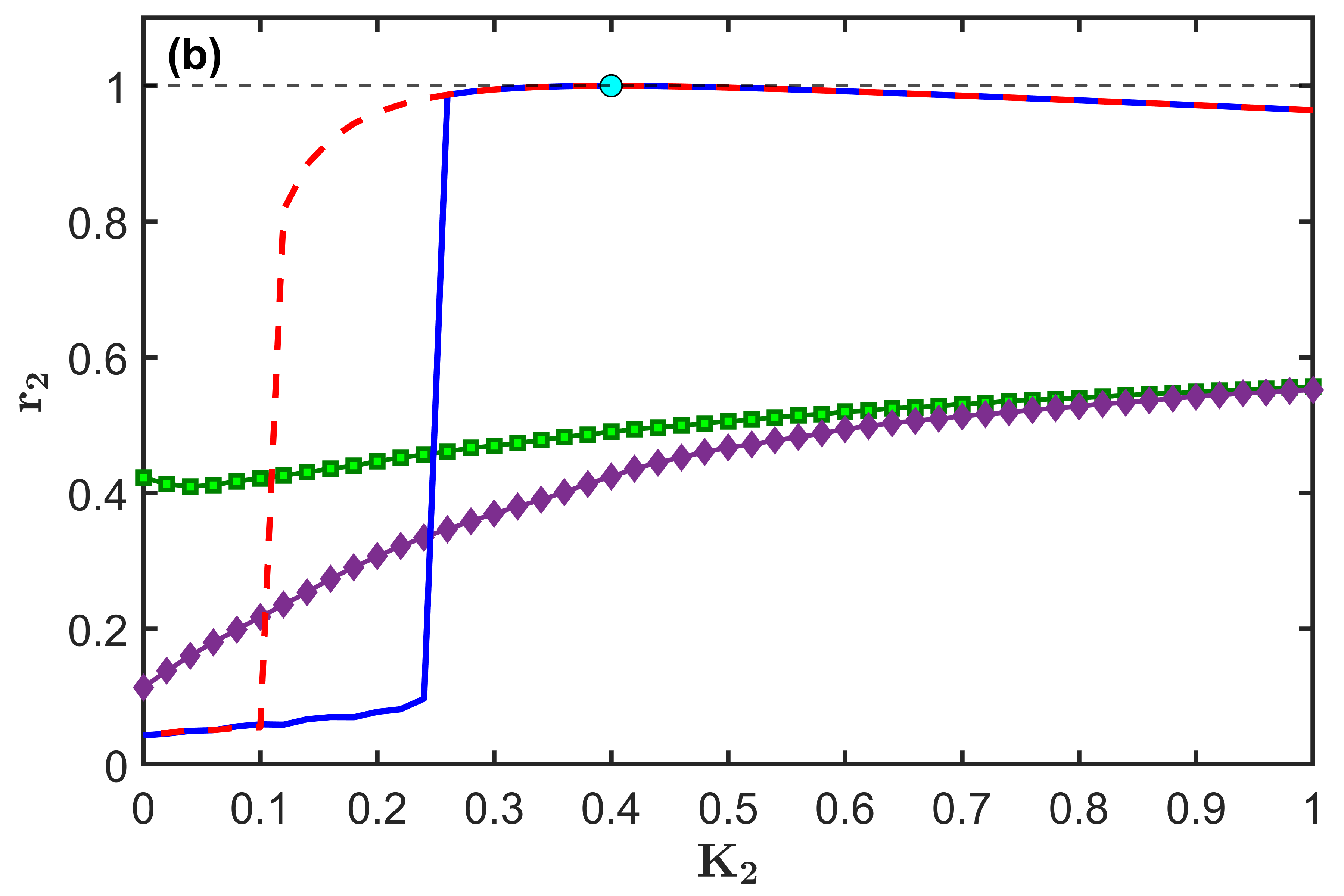}}

\subfigure{\includegraphics[width=0.45\textwidth]{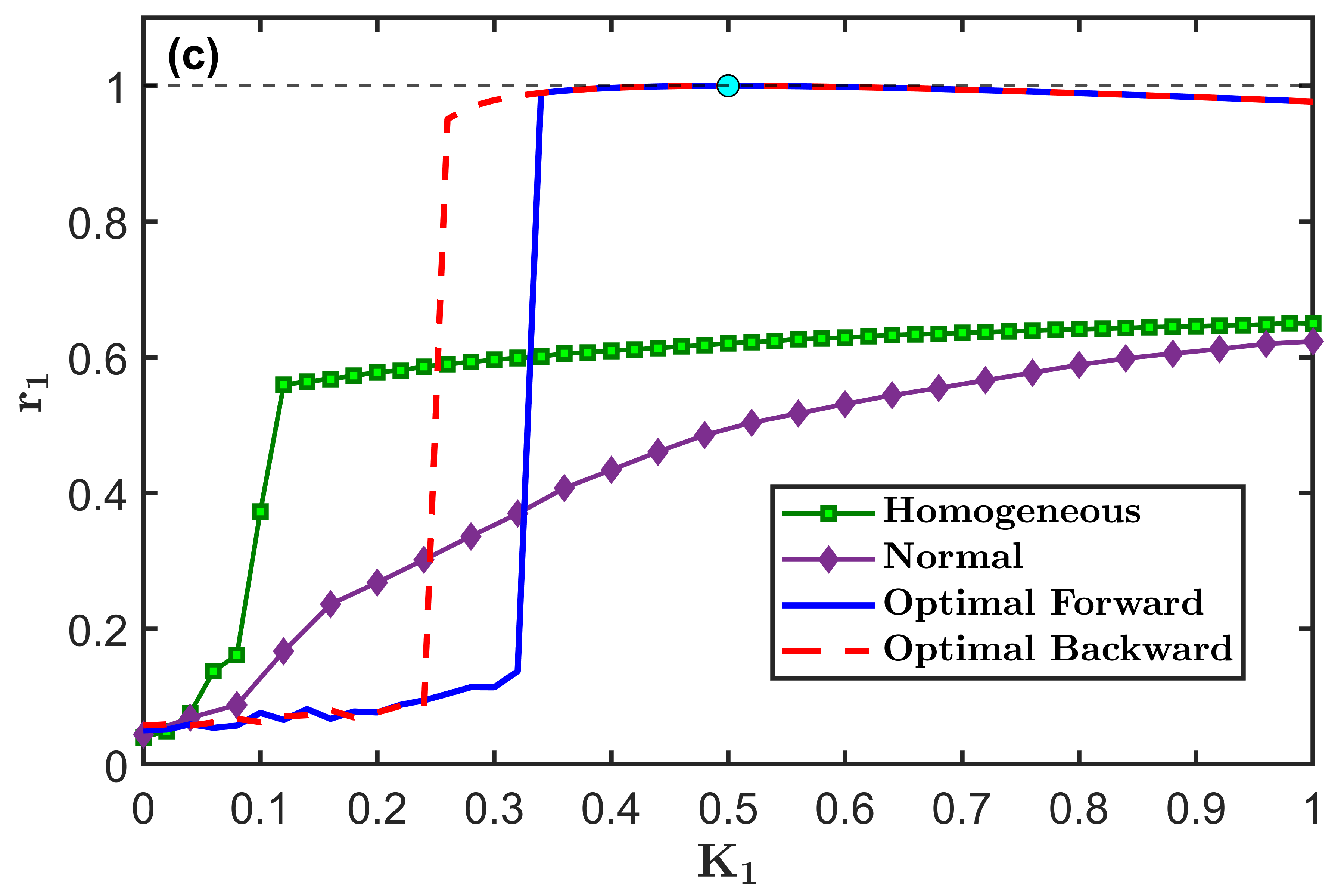}}
\subfigure{\includegraphics[width=0.45\textwidth]{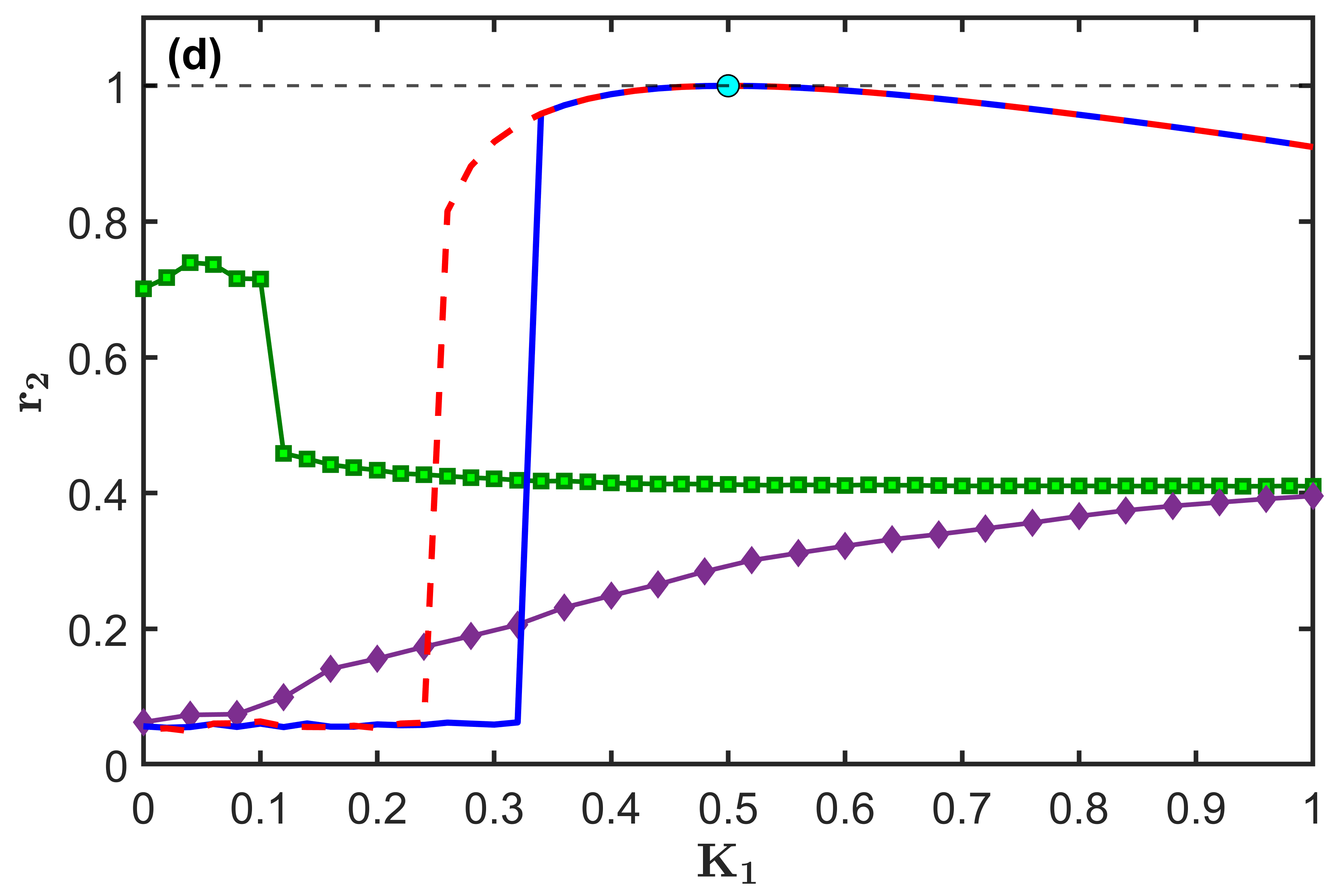}}
\caption{
\textbf{Synchronization dynamics in a scale-free network ($N = 500$, $\gamma = 2.8$) under phase frustration $\alpha = \beta = 0.5$.}
Purple diamond and green square markers denote the synchronization levels of the order parameters ($r_1$ and $r_2$) obtained from normal and homogeneous natural frequency distributions, respectively. 
The blue solid and red dashed curves correspond to forward and backward adiabatic continuation of the coupling strength, computed using the proposed optimal frequency assignment. 
\textbf{(a)--(b)} Variation of order parameters with the second-harmonic coupling $K_2$ at fixed $K_1 = 0.3$. 
\textbf{(c)--(d)} Variation of order parameters with the first-harmonic coupling $K_1$ at fixed $K_2 = 0.1$. 
In both parameter sweeps, the system undergoes a discontinuous (first-order) synchronization transition characterized by pronounced hysteresis. Complete synchronization is achieved at the prescribed target points (cyan markers), namely $K_2^{(p)} = 0.4$ in panels (a)–(b) and $K_1^{(p)} = 0.5$ in panels (c)–(d).
}
\label{fig:K_1_K_2_non_zero}
\end{figure*}


Now, we compare the real and imaginary parts of the

equation $r_2=r_2^{\text{lock}}+r_2^{\text{drift}}$. 
\begin{align}
    r_2 {\langle q \rangle}=\ &{\cos\beta} 
    \int_{q_{\min}}^\infty dq q P (q) \sqrt{1 - \left( \frac{2aq - \Omega}{2qK_2 r_2} \right)^2} \times \nonumber \\
    & H\left( 1 - \left| \frac{2aq - \Omega}{2qK_2 r_2} \right| \right) + \frac{\sin\beta}{2K_2r_2} \left(2a{\langle q \rangle}-\Omega\right)  \nonumber\\    &- \sin\beta\int_{q_{\min}}^\infty dq q P(q) \left( \frac{2aq - \Omega}{2qK_2 r_2} \right) \times \nonumber \\ & \sqrt{1 - \left( \frac{2qK_2 r_2}{2aq - \Omega} \right)^2}  H\left( \left| \frac{2aq - \Omega}{2qK_2 r_2} \right|-1 \right)
     \label{eq:real}
\end{align}

and

\begin{align}
  2a{\langle q \rangle}-\Omega= \ &\int_{q_{\min}}^\infty dq \left( {2aq - \Omega} \right) P(q) \sqrt{1 - \left( \frac{2qK_2 r_2}{2aq - \Omega} \right)^2} \nonumber \\& \times H\left(  \left| \frac{2aq - \Omega}{2qK_2 r_2} \right| -1\right) + 2K_2r_2\tan\beta \nonumber \\& \times \int_{q_{\min}}^\infty dq\, qP(q) \sqrt{1 - \left( \frac{2aq - \Omega}{2qK_2 r_2} \right)^2}  \nonumber \\& \times H\left(1-  \left| \frac{2aq - \Omega}{2qK_2 r_2} \right| \right)
  \label{eq:imaginary}
\end{align}
Using Taylor’s series expansion to equation  (\ref{eq:imaginary}),
\begin{align}
     2a{\langle q \rangle}-\Omega= \ &2K_2r_2\tan\beta \int_{\frac{\Omega}{2 \left( a+K_{2} r_{2}\right)}}^{\frac{\Omega}{2 \left( a-K_{2} r_{2}\right)}} dq\,qP(q) \nonumber \\& \times \sqrt{1 - \left( \frac{2aq - \Omega}{2qK_2 r_2} \right)^2} + \int_{q_{\min}}^\infty dq \,P(q) \nonumber \\& \times\left( {2aq - \Omega} \right) \biggl\{{1 - \frac{1}{2}\left( \frac{2qK_2 r_2}{2aq - \Omega} \right)^2} \biggl\}.
     \label{eq:Simplified_imaginary}
\end{align}
For more simplification take $ y=\frac{2aq - \Omega}{2K_2 r_2} $ and apply to the equation (\ref{eq:Simplified_imaginary}),
\begin{align}
    \ &\frac{2\tan\beta}{a}\int_{-\frac{\Omega}{2\left(a+K_2r_2\right)}}^\frac{\Omega}{2\left(a-K_2r_2\right)}dy\left(\frac{2K_2r_2y+\Omega}{2a}\right)P\left(\frac{2K_2r_2y+\Omega}{2a}\right)\nonumber\\& \times\sqrt{1-\frac{y^2}{\left(\frac{2K_2r_2y+\Omega}{2a}\right)^2}}-\int_{q_{\min}}^\infty dqP(q) \frac{\left(2q\right)^2}{2\left(2aq - \Omega\right)}=0
\end{align}
Taking $r \longrightarrow 0^+$ we get ,
\begin{align}
    \pi \left(\frac{\Omega_{c}}{2a}\right)^2 P\left(\frac{\Omega_c}{2a}\right) \tan\beta 
    &= \int_{q_{\min}}^\infty dq \, P(q) \left(\frac{q^2}{q - {\frac{\Omega_c}{2a}}}\right)
    \label{eq:Omega_critical},
\end{align}
where $\Omega_c$ represents the critical group angular velocity corresponding to the onset of synchronization.

From equation (\ref{eq:real}) and (\ref{eq:imaginary}) we get,
\begin{equation}
 r_2 {\langle q \rangle}={\frac{1}{\cos\beta}} 
   \int_{\frac{\Omega}{2 \left( a+K_{2} r_{2}\right)}}^{\frac{\Omega}{2 \left( a-K_{2} r_{2}\right)}} dq q P (q) \sqrt{1 - \left( \frac{2aq - \Omega}{2qK_2 r_2} \right)^2}
   \label{eq:r_2<q>}
\end{equation}
Following the earlier approach,  substituting $ y=\frac{2aq - \Omega}{2K_2 r_2} $ and from the equation (\ref{eq:r_2<q>}) we get,
\begin{align}
 {\langle q \rangle}=\ & \frac{K_2}{a\cos\beta}\int_{-\frac{\Omega}{2\left(a+K_2r_2\right)}}^\frac{\Omega}{2\left(a-K_2r_2\right)} dy \, P  \left(\frac{2K_2r_2y+\Omega}{2a}\right) 
 \nonumber \\& \times \left(\frac{2K_2r_2y+\Omega}{2a}\right)  \sqrt{1-\frac{y^2}{\left(\frac{2K_2r_2y+\Omega}{2a}\right)^2}}
 \label{eq:<q>}
\end{align}
Taking $r_{2} \longrightarrow 0^+$  and integrating the equation (\ref{eq:<q>}), the critical coupling $K_2^c$ for onset synchronization   ,
\begin{equation}
    K_2^c=\frac{8a^3\langle q \rangle\cos\beta}{\pi{\Omega_c}^2P\left(\frac{\Omega_c}{2a}\right)}
    \label{Critical_value}
\end{equation}
where $a = K^{(p)}_2 \sin\beta$ and the angular velocity $\Omega_c$ satisfy the equation (\ref{eq:Omega_critical}).

\begin{figure}
\centering

\includegraphics[width=\columnwidth]{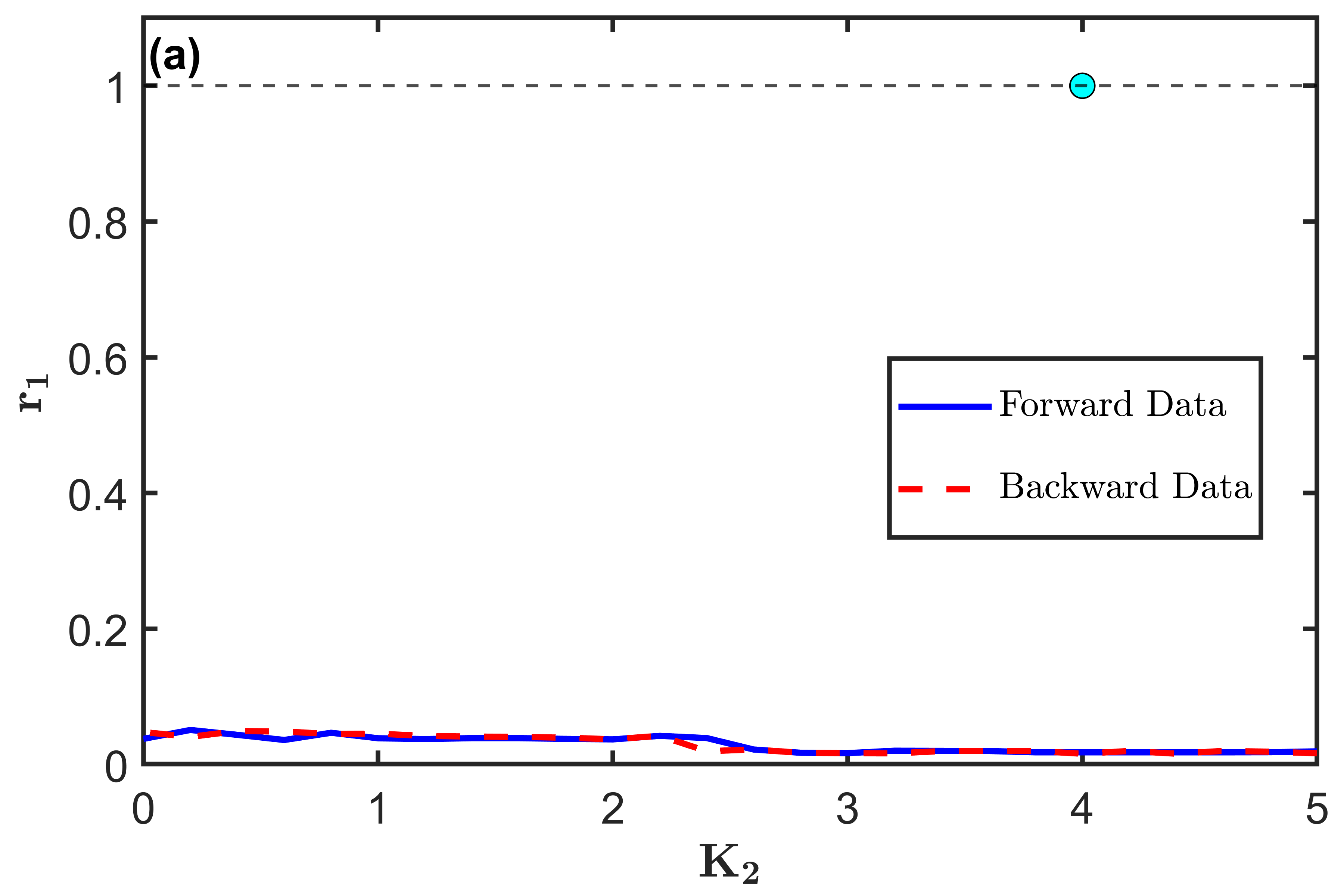}

\includegraphics[width=\columnwidth]{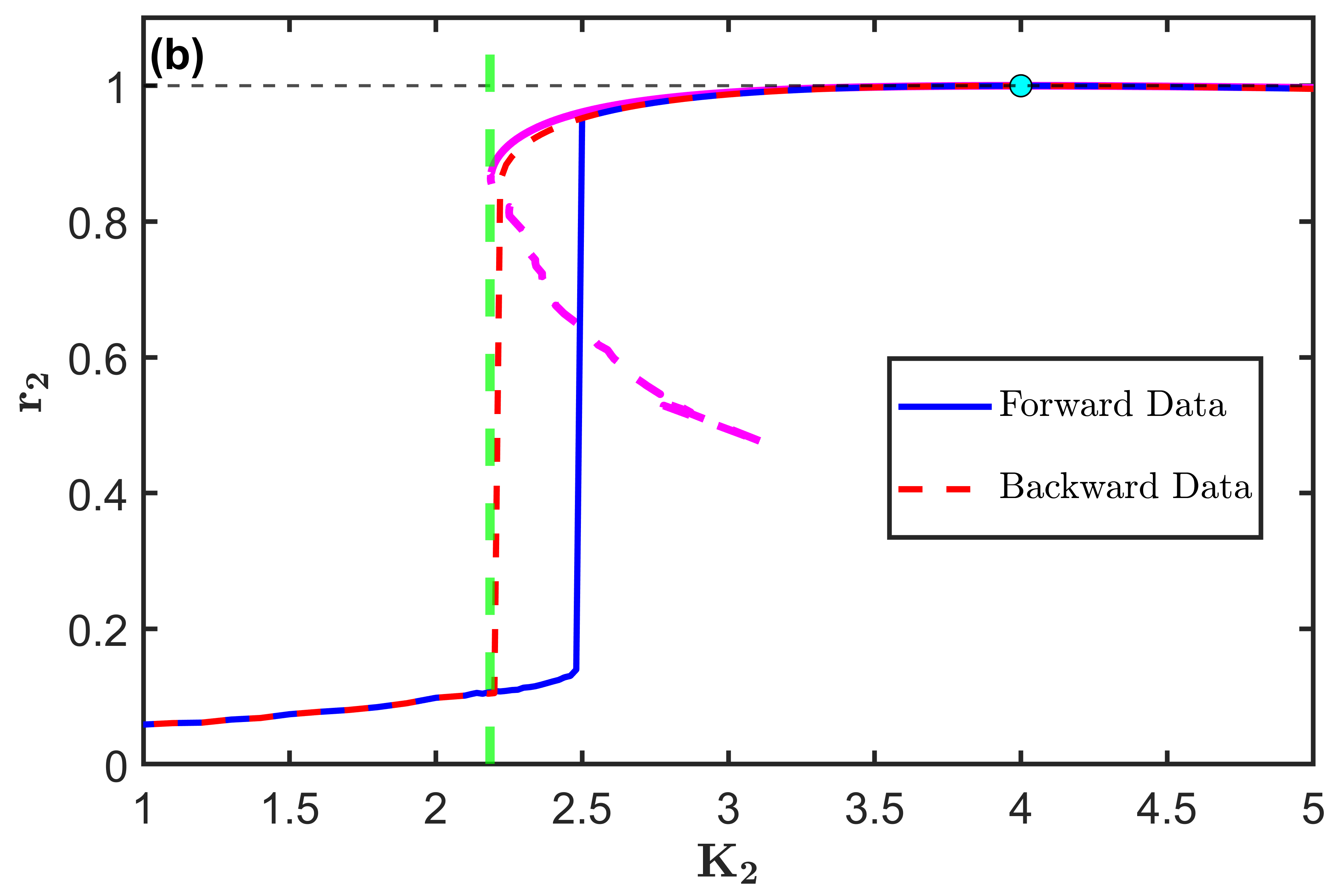}
\caption{\textbf{Order parameters $r_1$ and $r_2$ vs. coupling strength $K_2$.} Numerical and analytical results for a scale-free network ($N=500$, $\gamma=2.8$, $\langle q \rangle \approx 9$, $\beta=0.5$). Blue and red lines denote simulation results via forward and backward integration, respectively. \textbf{(a)} $r_1 \rightarrow 0$. \textbf{(b)} Bifurcation diagram of $r_2$ exhibiting a first-order transition and a distinct hysteresis area. Cyan dots at $K_2^{(p)}=4$ represent complete two cluster synchronization $(r_1 \rightarrow 0,r_2=1)$, provided $K_1=0$. Solid and dashed magenta curves indicate the stable and unstable analytical branches derived from Eqs. \eqref{eq:real} and \eqref{eq:imaginary}. The analytical threshold (green dashed line) $K^c_2 \approx 2.1846 < K^{(p)}_2=4$  from Eq. \eqref{Critical_value} shows agreement with the numerical simulation.}
\label{fig:K_1_0}
\end{figure}

\section{Results}\label{Results}
In this section, we validate the analytical derivation using frequency assignment derived in Eq.~(\ref{eq:optimal_frequency}) by confirming the emergence of complete synchronization at specific target points in the parameter space. We conduct numerical simulations of the Sakaguchi-Kuramoto model with bi-harmonic and higher-order coupling on different network topologies. The network ensemble includes a generated scale-free (SF) \cite{barabasi2003scale} network ($N = 500$, exponent $\gamma = 2.8$), Erdős–Rényi (ER) \cite{Erdos_PubMathDeb1959,Erdos_PMIHAS1960,Erdos_AMH1961} network $(N = 500)$, and two empirical networks often cited for their small-world properties: the Zachary Karate Club \cite{zachary1977information} ($N = 34$) and the \textit{C. elegans} neural network \cite{white1986structure} ($N = 131$).

The system dynamics are integrated using the fourth-order Runge-Kutta (RK4) method with a time step of $\delta t = 0.01$. Initial phase variables are drawn from a uniform random distribution, and transients are discarded to ensure the system reaches a steady state. For the following analysis, we fix the phase frustration parameters such that $\alpha = \beta$. It is important to note that while $\alpha$ and $\beta$ are chosen to be equal here for illustrative purposes, the optimal frequency set ensures complete synchronization at the target point even when $\alpha \neq \beta$. The numerical continuation is performed adiabatically in both forward and backward directions: the control parameter is incremented (or decremented) in small steps, with the final state of the previous run serving as the initial condition for the subsequent step. This procedure allows us to rigorously detect complete synchronization associated with discontinuous (or continuous) transitions.

We first investigate the synchronization dynamics in the SF network by setting a target coupling point at $K^{(p)}_{1}=0.5$ and $K^{(p)}_{2} = 0.4$.

In Figs.~\ref{fig:K_1_K_2_non_zero}(a) and \ref{fig:K_1_K_2_non_zero}(b), we fix the first coupling strength at $K_{1} = 0.3$ and vary $K_{2}$ in the range $[0, 1]$ with $\alpha = \beta = 0.5$. As illustrated, the system exhibits a clear first-order (discontinuous) phase transition. During the forward continuation (increasing $K_2$), the order parameters $r_1$ and $r_2$ remain low, indicating an incoherent state, until the coupling reaches a critical threshold. At this point, both parameters rise abruptly, converging asymptotically toward unity. Crucially, in the vicinity of the target value $K^{(p)}_{2} = 0.4$, the forward and backward continuation paths coincide. This implies that the system attains complete single cluster synchronization ($r_{1,2} \approx 1$) at $K^{(p)}_{2} = 0.4$ (marked by the cyan point) by following the same stable trajectory in both scanning directions.

\begin{figure}
\centering

\includegraphics[width=\columnwidth]{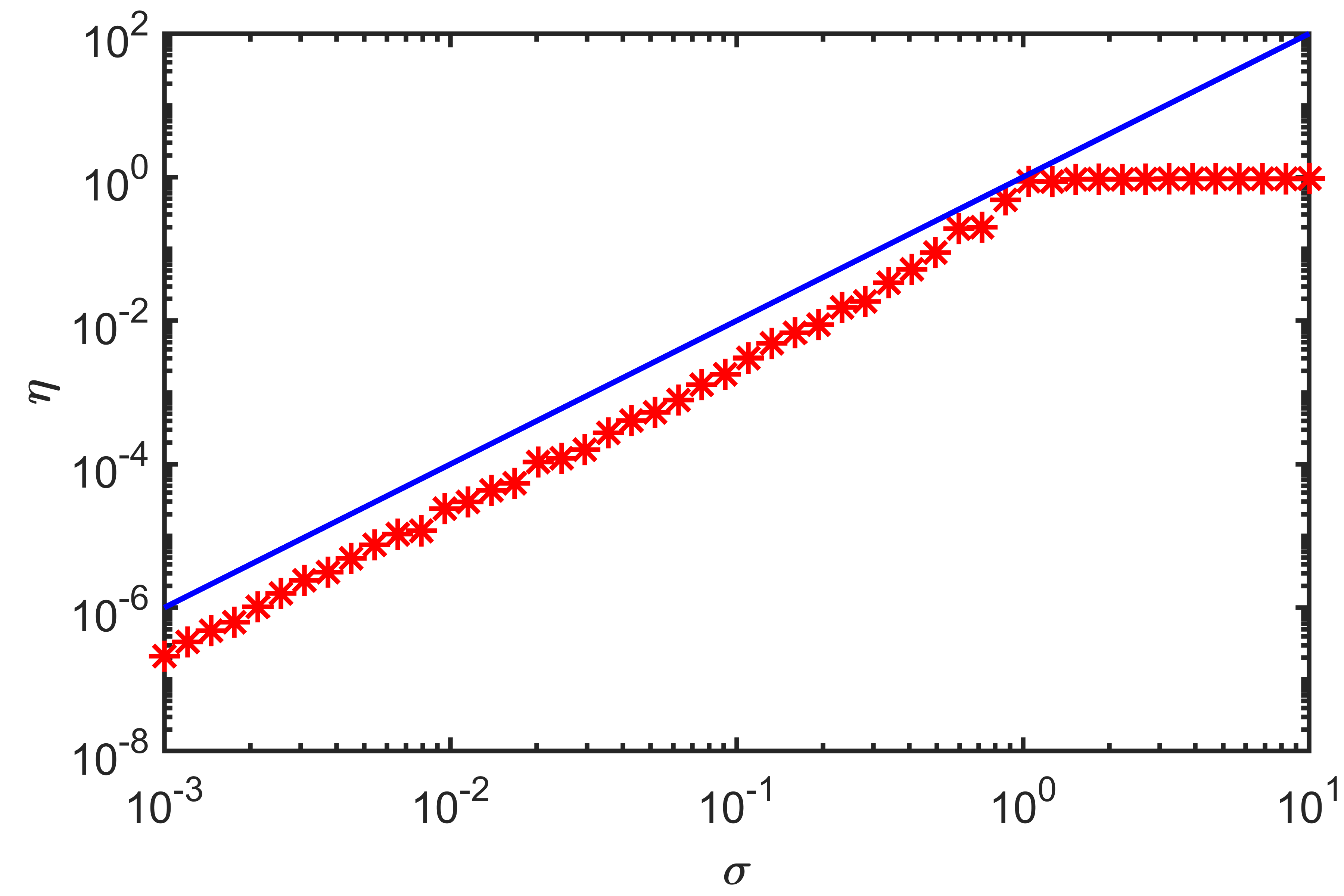}

\caption{\textbf{The effect of frequency deviation on system synchronization.} The graph compares the simulated error $\eta$ (red symbols) against the theoretical scaling limit $\eta \sim \sigma^2$ (blue line) under varying noise strengths $\sigma$. The system exhibits two distinct regimes: a quadratic growth of error at low $\sigma$, followed by a saturation phase ($\eta \longrightarrow  1$) indicating total loss of synchronization at high $\sigma$. Results are shown for a scale-free network 
($N = 500$, $\gamma = 2.8$); similar behavior is observed for other 
network types.}
\label{fig:noise}
\end{figure}

\begin{figure*}[t]
\centering

\subfigure{\includegraphics[width=0.45\textwidth]{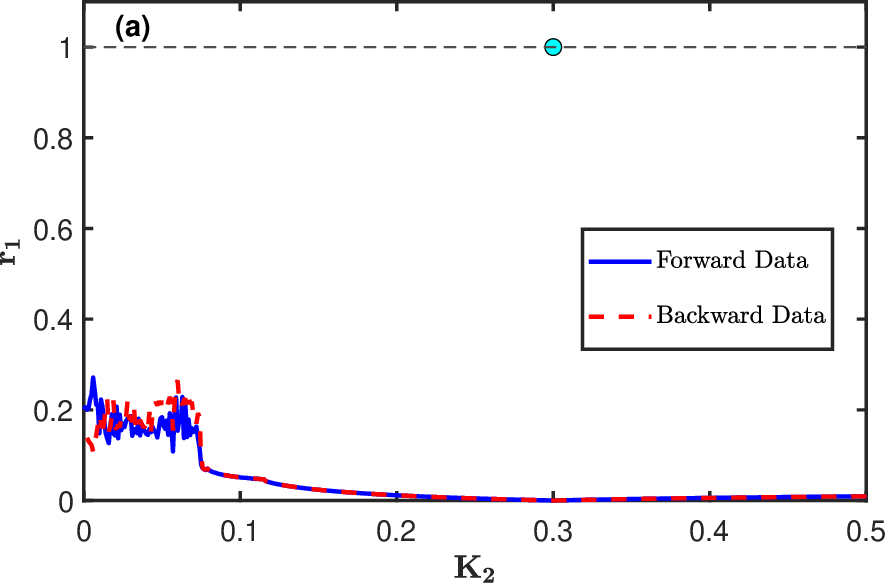}}
\subfigure{\includegraphics[width=0.45\textwidth]{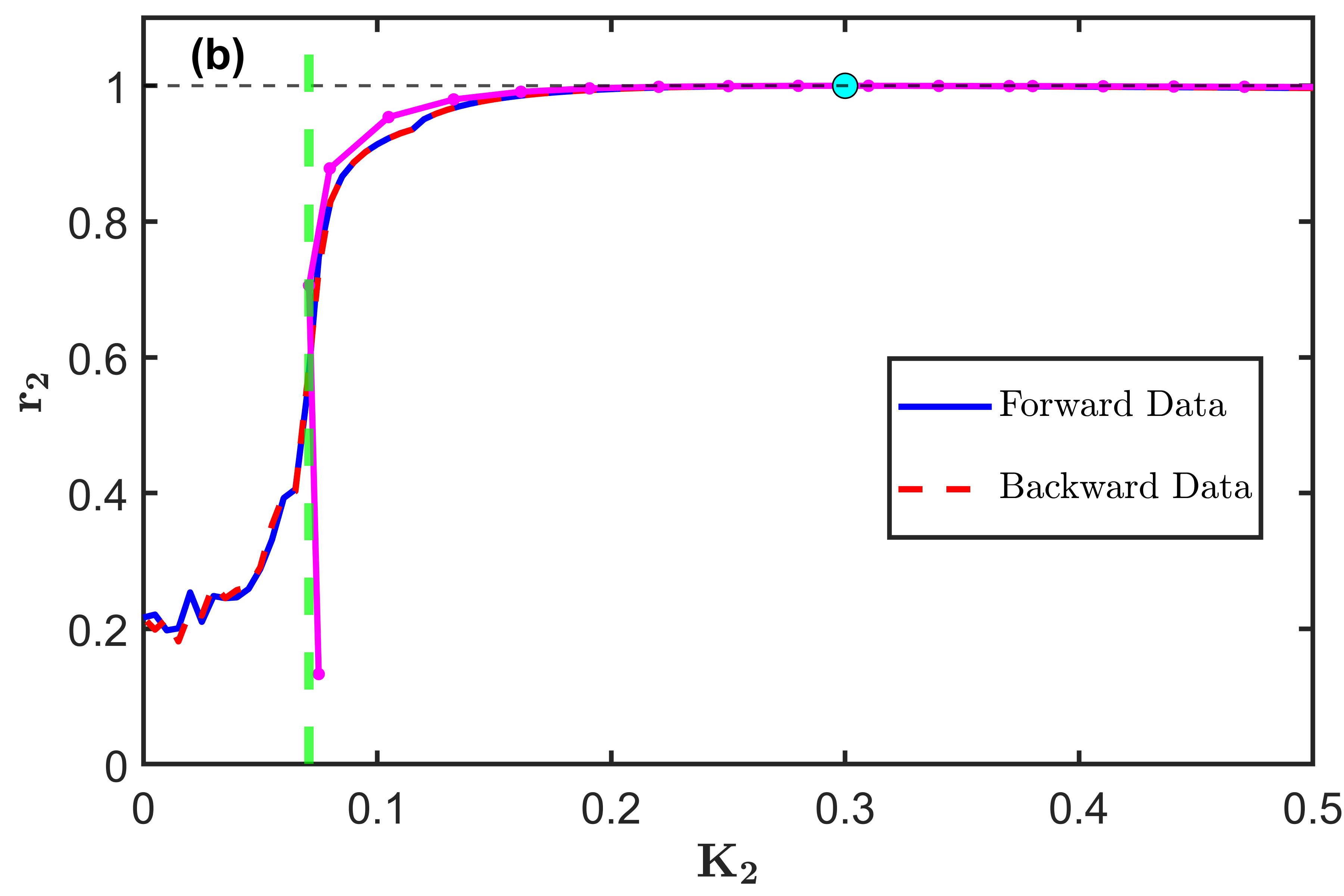}}

\subfigure{\includegraphics[width=0.45\textwidth]{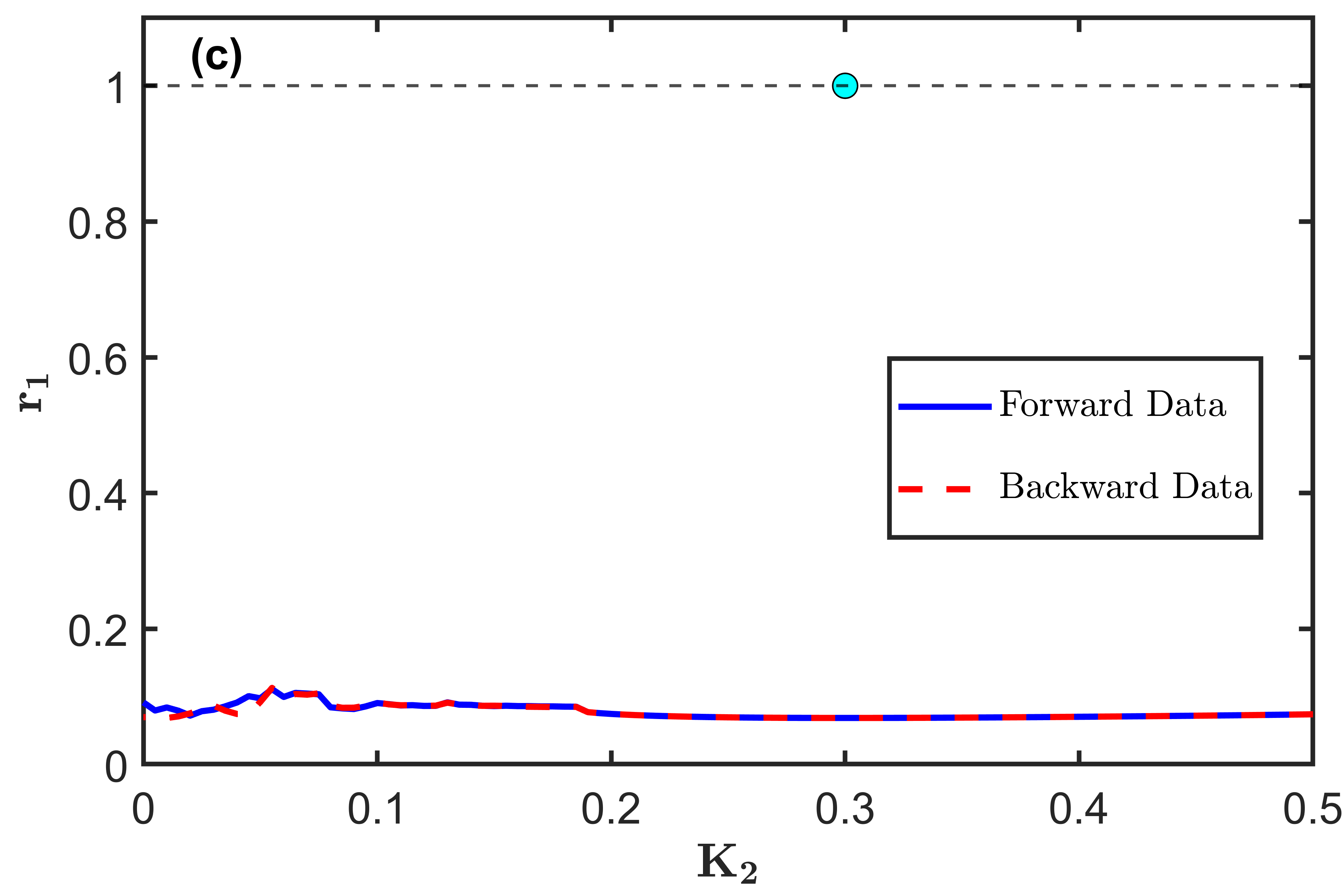}}
\subfigure{\includegraphics[width=0.45\textwidth]{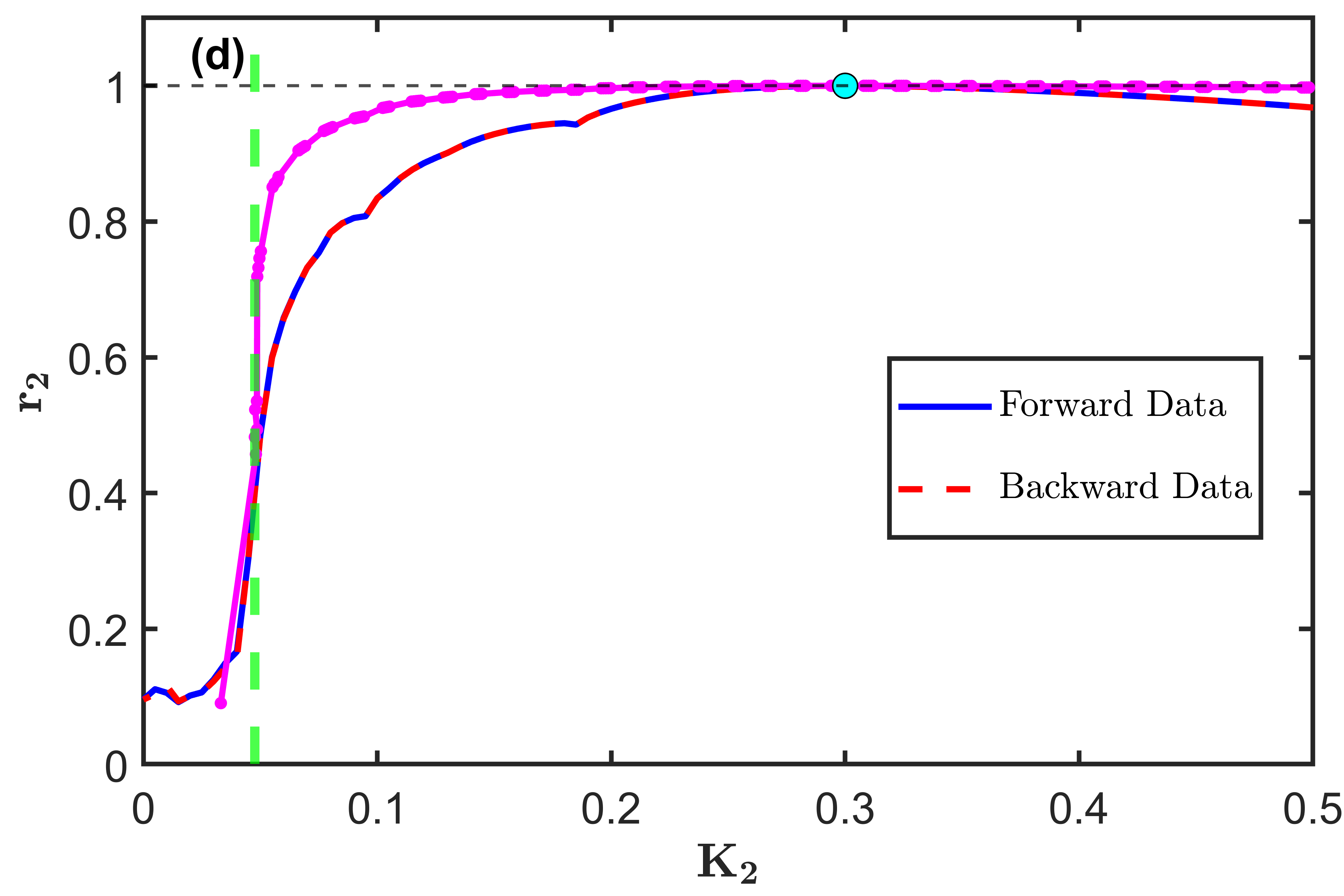}}

\caption{\textbf{Synchronization transitions in empirical networks.}
Variation of order parameters $r_1$ and $r_2$ as functions of the coupling strength $K_2$ for the Zachary Karate Club network ($N=34$) \textbf{(a,b)} and the \textit{C. elegans} network ($N=131$) \textbf{(c,d)} with $\beta = 0.2$.
Blue solid and red dashed curves denote forward and backward numerical simulations, respectively.
In both networks, $r_1 \rightarrow 0$ , whereas $r_2$ exhibits a reversible, continuous transition with overlapping forward and backward paths.
Magenta curves in \textbf{(b)} and \textbf{(d)} represent the analytical values of $r_2$ obtained from the self-consistency equations.
Cyan dots at $(0.3,1)$ indicate the target synchronization points.
The analytical critical couplings (green dashed lines), $K_2^{c} \approx 0.0708$ for the Karate Club and $K_2^{c} \approx 0.0476$ for \textit{C. elegans}, closely match the numerical estimates.}
\label{fig:Real_world}
\end{figure*}

Similarly, Figs.~\ref{fig:K_1_K_2_non_zero}(c) and \ref{fig:K_1_K_2_non_zero}(d) present the scenario where $K_2$ is fixed at the value $0.1$, and $K_1$ is varied in the range $[0, 1]$. Contrary to a continuous approach to synchronization, here we also observe significant hysteresis. The order parameters follow a distinct bistable path: remaining incoherent for low values of $K_1$ during the forward scan before jumping to a coherent state, while sustaining synchronization down to lower coupling values during the backward scan. However, specifically around the target point $K_1^{(p)} = 0.5$, the forward and backward branches merge. Both parameters reach exactly $1$ at the $K_1^{(p)} = 0.5$ (marked by the cyan point) along the same path in the vicinity of that point. Notably, due to the assignment of optimal natural frequencies, this complete synchronization state is preserved in both transition directions, regardless of whether $K_1$ is fixed while varying $K_2$, or vice versa.

In Fig.\ref{fig:K_1_K_2_non_zero}, the purple diamond and green square markers represent the synchronization levels of the order parameters obtained from the normal and homogeneous natural frequency distributions, respectively. Under the same conditions, the order parameters fail to reach complete synchronization and remain at comparatively lower values. This indicates that neither the normal nor the homogeneous frequency distributions can drive the system to complete synchronization under identical conditions, highlighting the effectiveness of the proposed optimal frequency distribution in enhancing synchronization.

\begin{figure*}[t]
\centering

\subfigure{\includegraphics[width=0.45\textwidth]{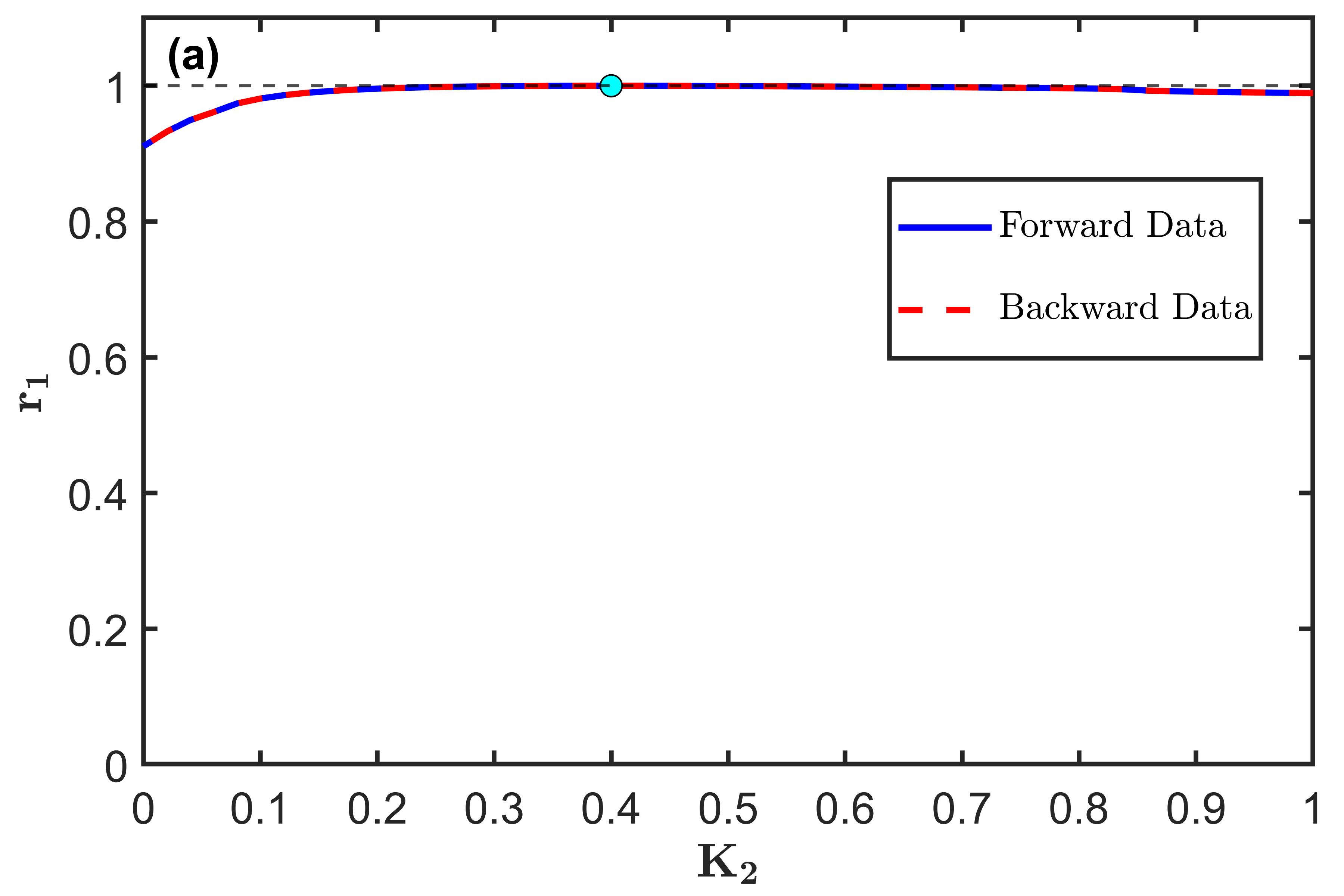}}
\subfigure{\includegraphics[width=0.45\textwidth]{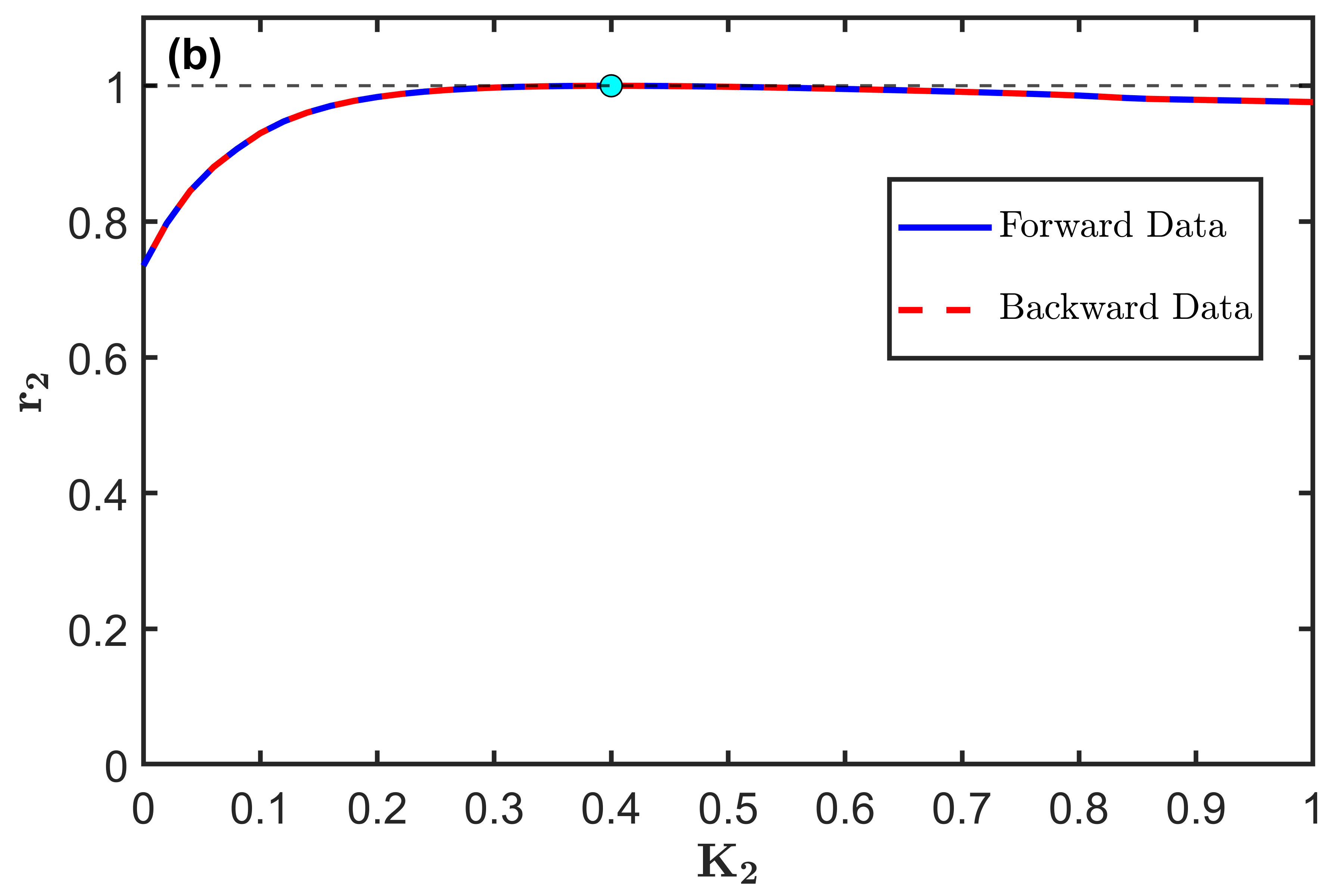}}

\subfigure{\includegraphics[width=0.45\textwidth]{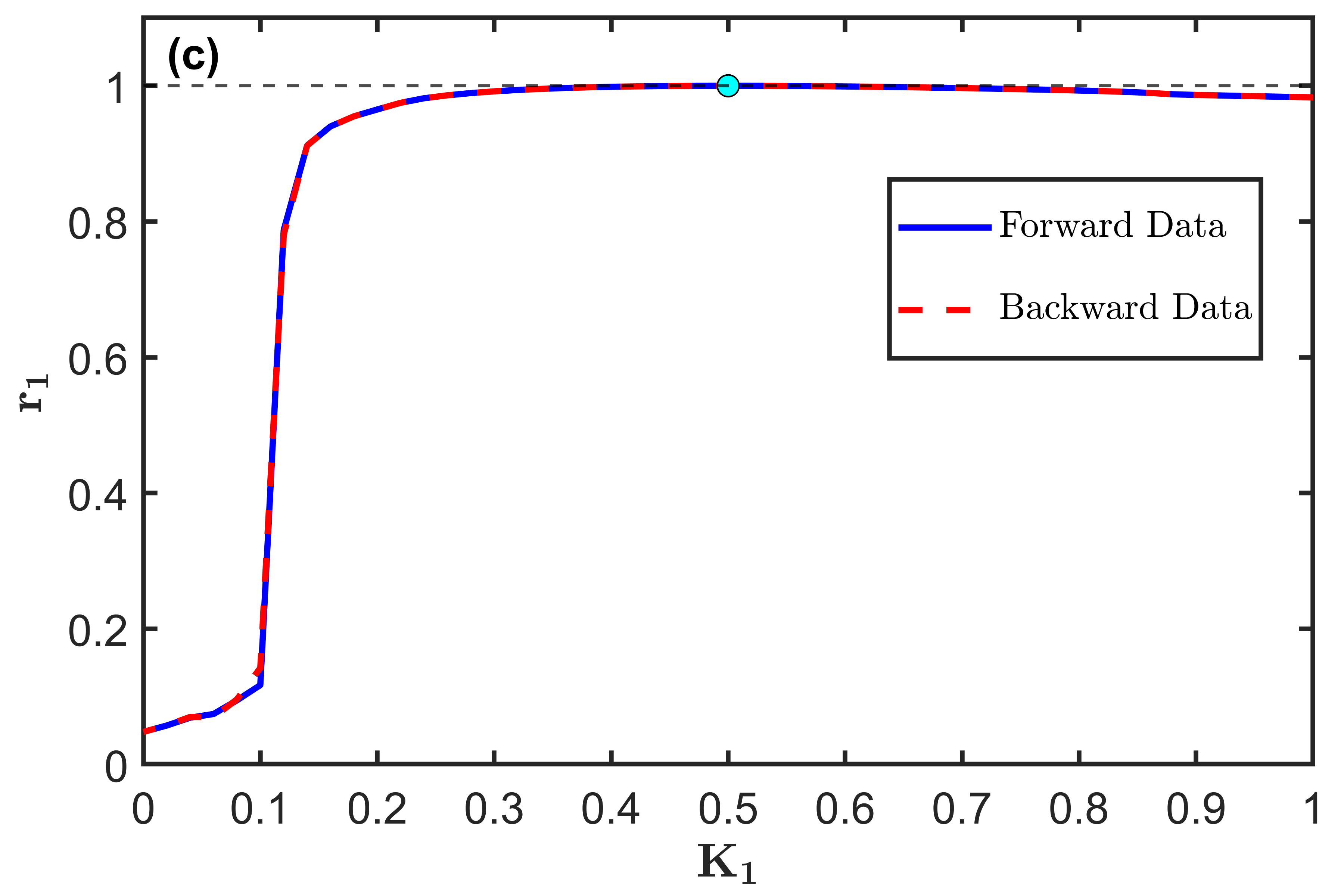}}
\subfigure{\includegraphics[width=0.45\textwidth]{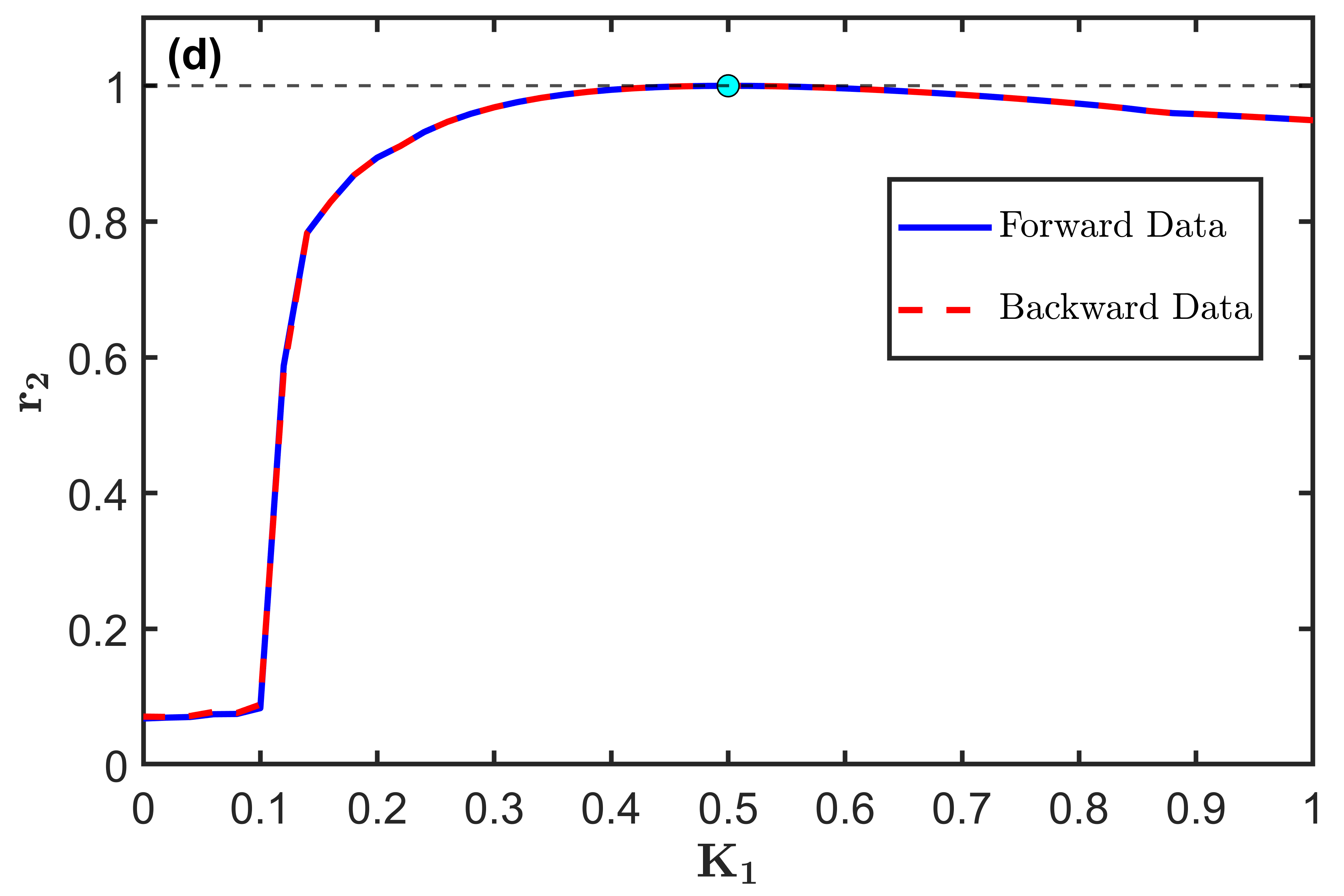}}
\caption{
\textbf{Synchronization dynamics in a Erdős–Rényi (ER) network with $N = 500$ and phase frustration $\alpha = \beta = 0.5$.} 
The blue solid and red dashed curves correspond to forward and backward adiabatic continuation of the coupling strength, computed using the proposed optimal frequency assignment. 
\textbf{(a)--(b)} Variation of order parameters with the second-harmonic coupling $K_2$ at fixed $K_1 = 0.3$. 
\textbf{(c)--(d)} Variation of order parameters with the first-harmonic coupling $K_1$ at fixed $K_2 = 0.1$. 
In both parameter sweeps, the system undergoes a second-order synchronization transition. Complete synchronization is achieved at the prescribed target points (cyan markers), namely $K_2^{(p)} = 0.4$ in panels (a)–(b) and $K_1^{(p)} = 0.5$ in panels (c)–(d).
}
\label{fig:ER_complete_syn}
\end{figure*}

In contrast, for the Erdős–Rényi (ER) network, the transition to complete synchronization occurs continuously, indicating a second-order synchronization transition. Unlike the scale-free case, no hysteresis is observed, and the order parameter increases smoothly with the coupling strength. This behavior is demonstrated in Fig. \ref{fig:ER_complete_syn} (Appendix).

To validate the mean field analytical results with the proposed optimal frequency distribution given by Eq.~(\ref{eq:optimal_K}), we conduct numerical simulations on Sakaguchi-Kuramoto model governed by the only second-harmonic coupling term in Eq.~(\ref{eq:2nd_harmonic}). Figures~\ref{fig:K_1_0}(a) and \ref{fig:K_1_0}(b) illustrate the evolution of the order parameters $r_1$ and $r_2$ as functions of the coupling strength $K_2$ for a scale-free network characterized by $N = 500$, $\gamma = 2.8$, and $\langle q \rangle \approx 9$. Throughout this analysis, the phase frustration parameter is held constant at $\beta = 0.5$.
The numerical results, obtained via forward (blue) and backward (red) integration, reveal distinct transition behaviors. Starting from an incoherent state and increasing $K_2$ adiabatically toward the target point $K^{(p)}_2 = 4$, we observe the path to synchronization. As shown in Fig.~\ref{fig:K_1_0}(a), the $r_1$ curve remains close to zero for any coupling strength.
The dynamics of the second order parameter, $r_2$, presented in Fig.~\ref{fig:K_1_0}(b). In this case, the system undergoes a clear first-order (discontinuous) phase transition. A significant hysteresis loop is observed, indicating a region of bistability. With $r_1=0$, the $r_2$ parameter reaches a state of complete two cluster synchronization ($r_2 = 1$) at the target point $K^{(p)}_2 = 4$, demonstrating the effectiveness of the optimal frequency distribution. Similar behavior is observed for the third-order harmonic, as presented in the Appendix.

These numerical findings are rigorously supported by our analytical framework. The bifurcation structure is derived from the self-consistency analysis in Eqs.~\eqref{eq:real} - \eqref{eq:imaginary}. Beyond the trivial incoherent solution ($r_2 = 0$), in Fig. \ref{fig:K_1_0}(b) the theory predicts two coherent branches: an unstable solution (dashed magenta curve), and a stable solution (solid magenta curve) that aligns with the  numerical simulation.
Crucially, the theoretical critical coupling strength, $K_c \approx 2.1846$, determined from Eq.~(\ref{Critical_value}), is represented by the green dashed line. The fact that this analytical threshold almost coincides with the numerical simulation confirms the validity of our theoretical approach. This high level of agreement demonstrates that the proposed optimal frequency distribution provides a predictable and controllable means of managing phase transitions in complex architectures.

The synchronization behavior of these coupled units when multiplicative noise is added to the natural frequency (in subsection \ref{Synchronization under the Presence of Noise}  of the analytical derivation) is illustrated in Fig.~\ref{fig:noise}, where increasing levels
of noise are introduced into the optimal frequency set. For small noise
amplitudes, the numerical results  confirm the quadratic scaling
$\eta \sim \sigma^2$ (solid blue line) in Eq. \eqref{eq:noise}. In the large-noise limit
$\sigma \to 1$, the natural frequencies become dominated by noise,
synchronization lost ($r \to 0$), and consequently $\eta \to 1$.

\subsection*{Verification on Real-World Networks}
To demonstrate the robustness of the proposed theoretical framework, we extend our analysis to two empirical network topologies: the Zachary Karate Club network ($N = 34$) and the \textit{C. elegans} neural network ($N = 131$). As with the previous analysis of scale-free networks, we evaluate the synchronization dynamics under the influence of the optimal frequency distribution. For both empirical cases, the phase frustration parameter $\beta$ is set to $0.2$.
Fig.~\ref{fig:Real_world} presents the variation of order parameters $r_1$ and $r_2$ as functions of the coupling strength $K_2$. For the Karate Club network, Fig.~\ref{fig:Real_world}(a) illustrates that $r_1$ tends to zero for any coupling value. Interestingly, as shown in Fig.~\ref{fig:Real_world}(b), the forward and backward simulation paths of $r_2$ follow the same trajectory, indicating the absence of a hysteresis loop in this specific topology. Despite the structural differences from the scale-free model, the network reaches complete two-cluster synchronized state at the targeted point $K_2^{(p)} = 0.3$ (marked by cyan dots). The analytical prediction for $r_2$ (magenta curve) yields a critical coupling strength of $K_2^c \approx 0.0708$ (green dashed line), which is found to be close to the numerical simulation.

Similar results are observed for the \textit{C. elegans} network in Figs.~\ref{fig:Real_world}(c) and \ref{fig:Real_world}(d), where the system reaches complete two-cluster synchronization at $K_2^{(p)} = 0.3$. The theoretical framework predicts a critical coupling strength of $K_2^c \approx 0.0476$ (green dashed line). While the analytical curve and the simulation results show minor deviations due to the finite size and inherent irregularities of the empirical connections, the values remain close to the predicted thresholds.

Furthermore, (in Appendix) we extend our results to the model only with third-order harmonic (Fig. \ref{third-order harmonic}) and generalized SK model of $N$ coupled oscillators (nodes) with $m$ - harmonics  ($m \geq 2$) \cite{skardal2011cluster} in \eqref{Higher order harmonic}.
The consistent performance across ER, SF networks and along with the empirical datasets demonstrates the generalizability of the proposed optimal frequency distribution. Also, the fact that the analytical curve and critical coupling $K_c$ remain in close alignment with simulation results, highlights the efficacy of the analytical approach in capturing the synchronization dynamics in diverse complex systems. 

\section{Conclusions}\label{Conclusions}

In this study, we proposed the complete single (or two) cluster synchronization of coupled bi-harmonic (or pure second-harmonic) Sakaguchi-Kuramoto oscillators. We have formulated a set of optimal natural frequencies that is robust against added noise, confirming complete (single or two-cluster) synchronization at some targeted, significantly weak coupling strength. Here, we introduced the synchrony alignment function (SAF) to find the optimal frequency that depends on heterogeneous network structure (adjacency matrix, $A= (A_{ij})_{N \times N}$), coupling strengths, and phase-frustrated coupling function. We use different types of networks with the proposed optimal set of frequencies; in all cases, we observe complete synchronization (either single or two clusters) at targeted coupling strength, while no other frequency can achieve complete synchronization, even with a higher coupling value. Next, by using mean field theory, we find analytically critical coupling strength in the presence of optimal frequency for pure second-harmonic SK oscillators. Here, we find a perfectly ordered symmetric two-cluster synchronized state at the targeted coupling strength. The results of numerical and analytical simulations agree well with each other. At last, we generalize results for higher harmonic oscillators.



\section{Acknowledgments}
The author gratefully acknowledges the Council of Scientific and Industrial Research (CSIR), New Delhi, India, for financial support under the CSIR-SRF (NET) fellowship program (File No.: 31/0019(13167)/2022-EMR-I).

\section{Declaration of generative AI and AI-assisted technologies in the writing process}
During the preparation of this work, the authors used language refinement tools in order to refine the english style of the presentation alongside the grammatical corrections. After using this tool/service, the authors reviewed and edited the content as needed and take full responsibility for the content of the publication.

\bibliographystyle{cas-model2-names}

\bibliography{cas-refs}



\begin{figure*}[t]
\centering
\subfigure{\includegraphics[width=0.45\textwidth]{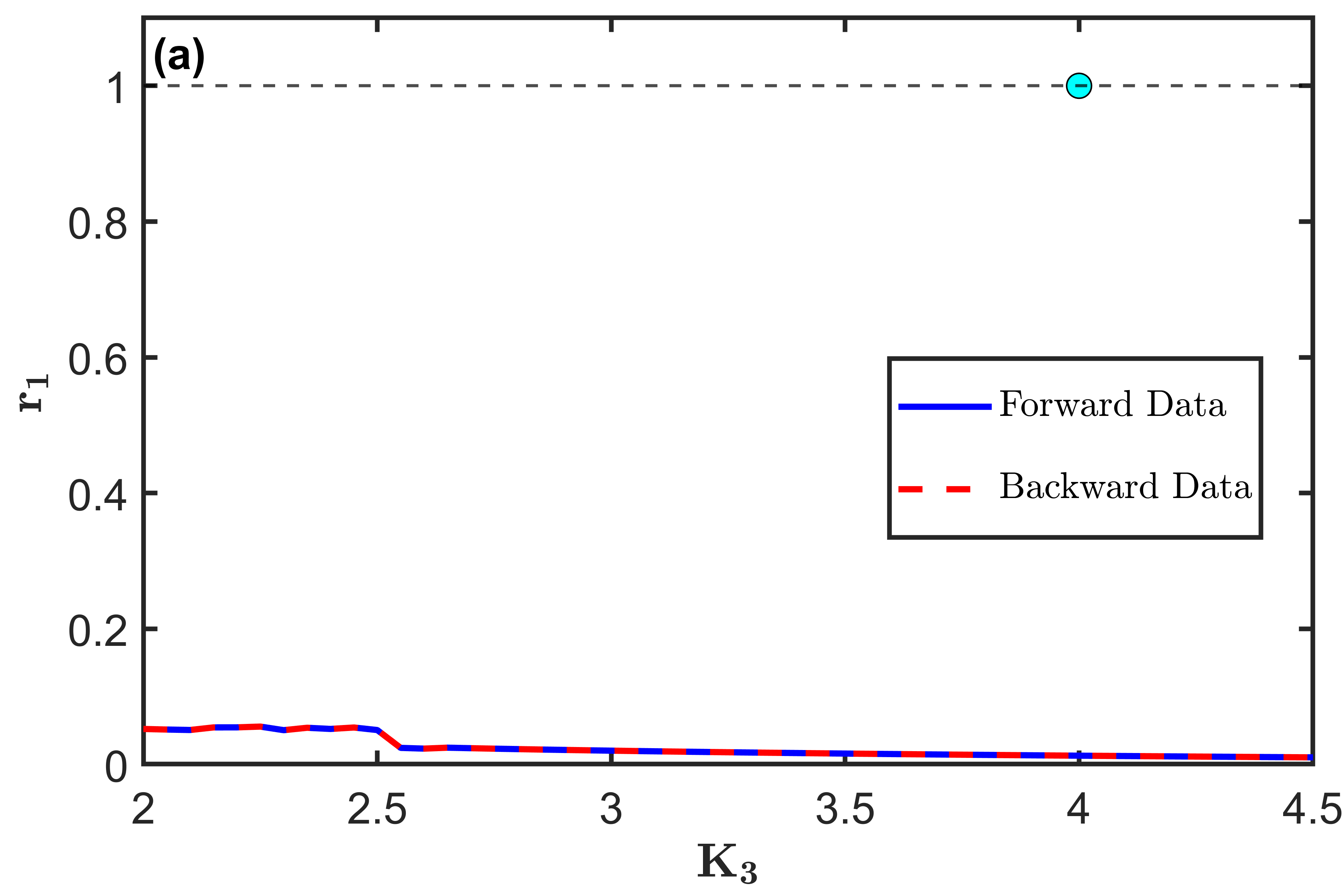}}
\subfigure{\includegraphics[width=0.45\textwidth]{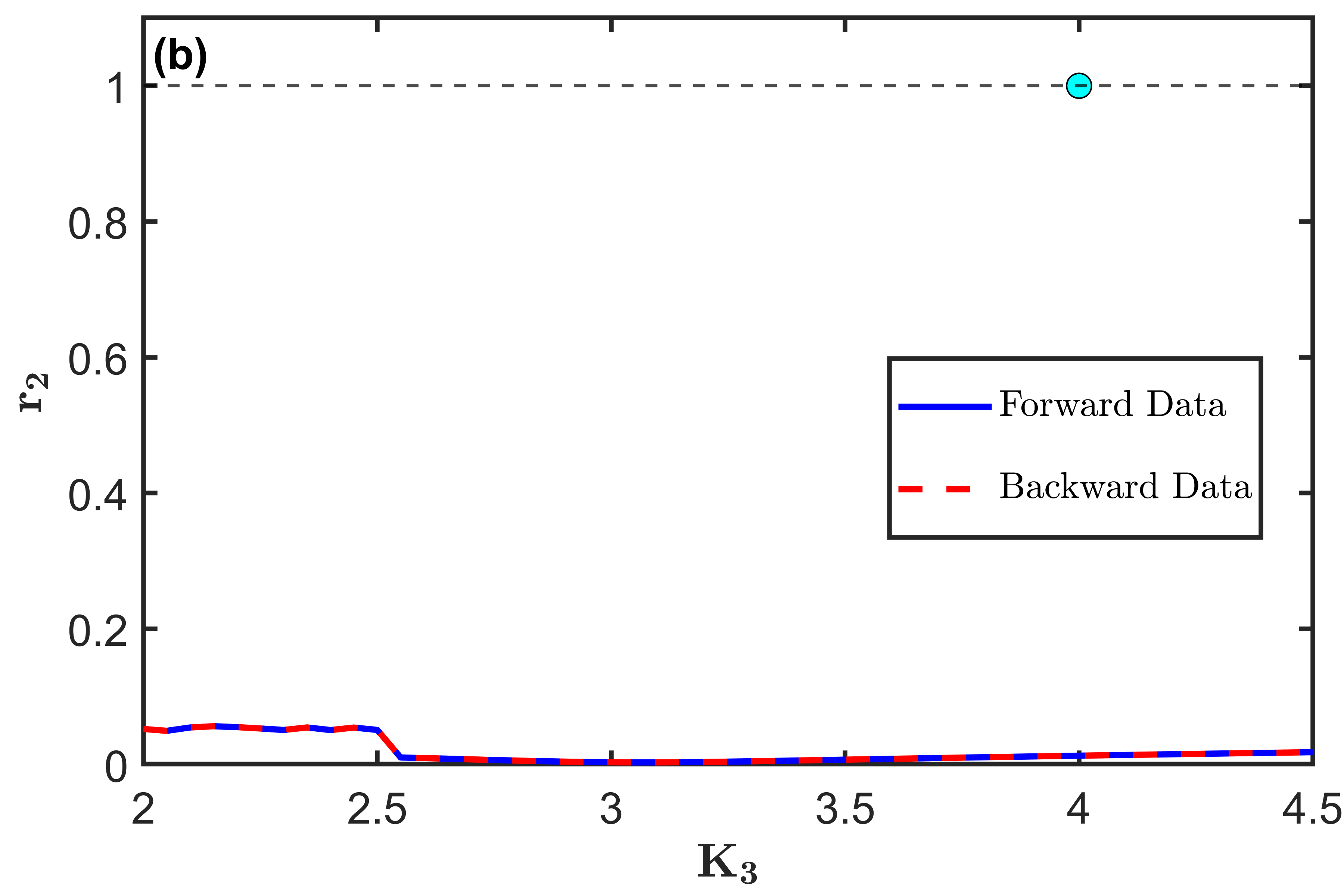}}
\subfigure{\includegraphics[width=0.45\textwidth]{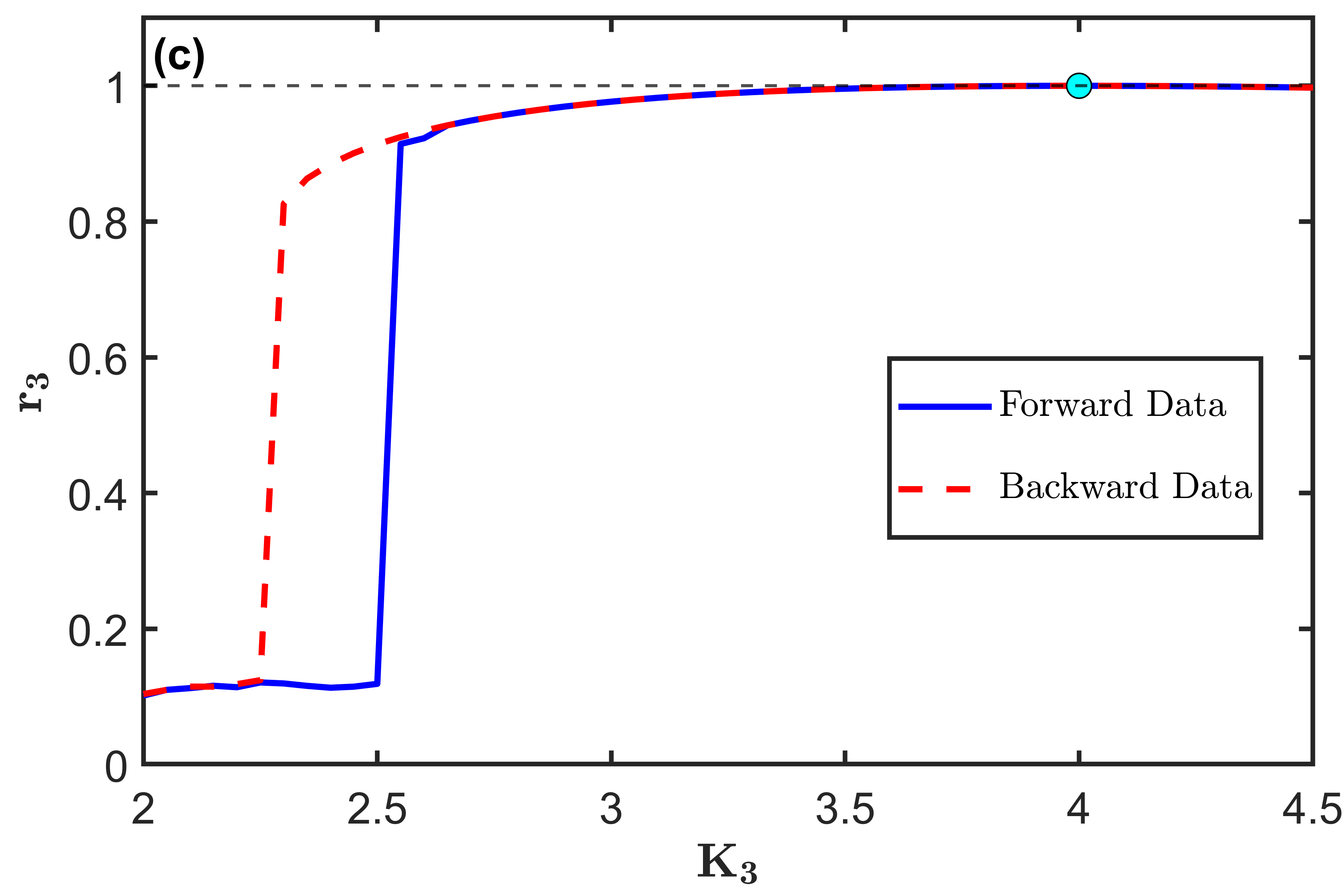}}

\caption{\textbf{Pure third-order harmonic extension of the coupled oscillator model.}
Order parameters $r_1$, $r_2$, and $r_3$ as functions of the coupling strength for a scale-free network with degree exponent $\gamma = 2.8$ and size $N = 500$, where phase frustration $0.5$.
Blue and red curves represent forward and backward simulations obtained using the proposed optimal frequency distribution.
The maximum value of $r_3$ appears at the targeted point $(4,1)$ (cyan dot) when $K_1 = 0$ and $K_2 = 0$.
While $r_1$ and $r_2$ tending to zero, $r_3=1$ exhibits complete multi-cluster synchronized state, consistent with the behavior observed above previous results.}
\label{third-order harmonic}
\end{figure*}

\newpage

\section*{Appendix}
Consider the generalized SK model of $N$ coupled oscillators (nodes) with $m$ - harmonics \cite{skardal2011cluster},
 \begin{eqnarray}
\frac{d\theta_i}{dt} &=& \omega_i + \sum_{j=1}^{N} \sum_{l=1}^m{K}_{l}A_{ij}{F}(l(\theta_j - \theta_i)-\alpha_{ij}^{(l)})
\label{Higher order harmonic}
\end{eqnarray}
where $\omega_i$ represents the natural frequency of node $i$, $A=(A_{ij})_{N \times N}$ is an adjacency matrix that determines the underlying interaction network,
$\mathrm{K}_l$ is the coupling strength for the $l^{th}$ harmonic , and the functional form of the coupling is captured by
${F}(l(\theta_j - \theta_i)-\alpha_{ij}^{(l)})=sin({l(\theta_j - \theta_i)-\alpha_{ij}^{(l)}})$ with distributed phase  frustration of $\alpha_{ij}^{(l)}$ that signifies the existence of frustration in each coupling function.

From Eq. (\ref{eq:J(w,L)}) , we get the general form of optimal frequency for complete $m$-cluster synchronization as, 
\begin{align}
    J(\bm{\omega},\bm{L})= \ & 0 \implies {{\bm{\omega}}} +   {\bm{{\tilde{d}}}}=0 \nonumber \\& \implies \bm{\omega}= -\sum_{j=1}^{N} A_{ij} [K_{1}F(-\alpha_{ij}^{(1)}) \nonumber \\& +K_{2}F(-\alpha_{ij}^{(2)})+... + K_{m}F(-\alpha_{ij}^{(m)})] 
\end{align}

Therefore, the compact form of the optimal frequency for complete $m$-cluster synchronization,  
\begin{equation}
    \bm{\omega}= \sum_{j=1}^{N} \sum_{l=1}^m\mathrm{K}_{l} A_{ij} Sin(\alpha_{ij}^{(l)}) 
    \label{Gen. optimal frequency}
\end{equation}
If we extend the calculation of mean field analysis for $N$ oscillators with pure $m^{th}$ order harmonic (when  $K_1=K_2=...=K_{(m-1)}=0$ , $K_{m} \neq 0$, and all $\alpha_{ij}^{(m)}$ are identical, say $\zeta$ ), then the general form of critical coupling (extension form of Eq. \eqref{Critical_value}) for the $m^{th}$ order harmonic is given by, 
\begin{equation}
    K_m^c=\frac{2a\langle q \rangle\cos\zeta}{\pi\left({\frac{\Omega_c}{ma}} \right)^2P\left(\frac{\Omega_c}{ma}\right)}
\end{equation}
along with the general form of Eq. \eqref{eq:Omega_critical} ,
\begin{equation}
    \pi \left(\frac{\Omega_{c}}{ma}\right)^2 P\left(\frac{\Omega_c}{ma}\right) \tan\zeta 
    = \int_{q_{\min}}^\infty dq \, P(q) \left(\frac{q^2}{q - {\frac{\Omega_c}{ma}}}\right)
\end{equation}
\subsection*{Complete synchronization for third-order harmonic } 
We extend the analysis to the third-order harmonic case $(m=3 ~\text {and} ~K_1=K_2=0,K_3\neq 0)$ with optimal frequency \eqref{Gen. optimal frequency} in order to examine the higher-harmonic synchronization behavior of the system \eqref{Higher order harmonic}. Fig.~\ref{third-order harmonic} shows the variation of the order parameters \(r_1\), \(r_2\), and \(r_3\) as functions of the coupling strength for a scale-free network with degree exponent \(\gamma = 2.8\), network size \(N = 500\), and  phase  frustration parameters fixed at $0.5$. The blue and red curves correspond to the results obtained from forward and backward simulations, respectively, using the proposed optimal frequency distribution.
In Fig.~\ref{third-order harmonic}, $r_1,r_2 \rightarrow 0$ along with $r_3=1$ demonstrate complete multi-cluster synchronized state. In contrast, the \(r_3\) curve [Fig.~\ref{third-order harmonic}(c)] displays a pronounced hysteresis loop, suggesting the coexistence of multiple stable states and a clear first-order phase transition for the third harmonic mode.
The qualitative behavior observed in the third-order harmonic case is consistent with the results obtained for the second-harmonic scenario, indicating that the underlying synchronization dynamics follow a similar pattern across different harmonic orders. This observation supports the generalization of the proposed theoretical framework to an arbitrary \(m\)-harmonic formulation, thereby extending the applicability of our approach to a broader class of coupled oscillator systems.












\end{document}